\documentclass[acmsmall,nonacm,screen]{acmart}  

\usepackage{algorithm}
\usepackage[noend]{algpseudocode}
\usepackage{alltt}
\usepackage{booktabs}
\usepackage[commandnameprefix=ifneeded,final]{changes}  
\usepackage{enumitem}
\usepackage{fontawesome5}
\usepackage{multirow}
\usepackage[autolanguage]{numprint} \npthousandsep{,}
\usepackage{soul}
\usepackage{subcaption}
\usepackage{textcomp}

\graphicspath{{figs/}{./}}

\renewcommand{\Pr}[1]{\ensuremath{\mathrm{Pr}\left[#1\right]}}
\newcommand{\E}[1]{\ensuremath{\mathrm{E}\left[#1\right]}}
\newcommand{\true}{\textsc{true}}
\newcommand{\false}{\textsc{false}}
\newcommand{\correcticon}{\faIcon{check-circle}}
\newcommand{\secureicon}{\faIcon{lock}}
\newcommand{\timeicon}{\faIcon{tachometer-alt}}
\newcommand{\memoryicon}{\faIcon{database}}
\newcommand{\cve}[1]{\href{https://www.cve.org/CVERecord?id=#1}{#1}}

\makeatletter
\algrenewcommand\ALG@beginalgorithmic{\small}
\algrenewcommand\alglinenumber[1]{\footnotesize #1:}
\makeatother

\newtheorem{recommend}{Recommendation}

\makeatletter
\newif\ifanon
\@ifclasswith{acmart}{anonymous}{\anontrue}{\anonfalse}
\newif\ifnonacm
\@ifclasswith{acmart}{nonacm}{\nonacmtrue}{\nonacmfalse}
\makeatother

\newif\ifcomment
\commenttrue    

\newif\iffigabbrv
\figabbrvtrue    
\newcommand{\figtext}{\iffigabbrv Fig.\else Figure\fi}
\newcommand{\figstext}{\iffigabbrv Figs.\else Figures\fi}

\newif\ifeqabbrv
\eqabbrvtrue    
\newcommand{\eqtext}{\ifeqabbrv Eq.\else Equation\fi}
\newcommand{\eqstext}{\ifeqabbrv Eqs.\else Equations\fi}

\title{A Taxonomy and Comparative Analysis of IPv4 Identifier Selection Correctness, Security, and Performance}

\author{Joshua J. Daymude}
\orcid{https://orcid.org/0000-0001-7294-5626}
\affiliation{%
    \institution{Arizona State University}
    \department{Biodesign Center for Biocomputing, Security and Society}
    \department{School of Computing and Augmented Intelligence}
    \streetaddress{727 E. Tyler St.}
    \city{Tempe}
    \state{AZ}
    \postcode{85281}
    \country{USA}}
\email{jdaymude@asu.edu}

\author{Antonio M. Espinoza}
\orcid{https://orcid.org/0009-0003-3175-2029}
\affiliation{%
    \institution{Eastern Washington University}
    \department{College of Science, Technology, Engineering, and Mathematics}
    \streetaddress{601 E Riverside Ave.}
    \city{Spokane}
    \state{WA}
    \postcode{99202}
    \country{USA}}
\email{aespinoza17@ewu.edu}

\author{Holly Bergen}
\orcid{https://orcid.org/0009-0004-3570-5120}
\affiliation{%
    \institution{Arizona State University}
    \department{Biodesign Center for Biocomputing, Security and Society}
    \department{School of Computing and Augmented Intelligence}
    \streetaddress{727 E. Tyler St.}
    \city{Tempe}
    \state{AZ}
    \postcode{85281}
    \country{USA}}
\email{sbergen2@asu.edu}

\author{Benjamin Mixon--Baca}
\orcid{https://orcid.org/0000-0002-4670-4578}
\affiliation{%
    \institution{Arizona State University}
    \department{Biodesign Center for Biocomputing, Security and Society}
    \department{School of Computing and Augmented Intelligence}
    \streetaddress{727 E. Tyler St.}
    \city{Tempe}
    \state{AZ}
    \postcode{85281}
    \country{USA}}
\email{bmixonba@asu.edu}

\author{Jeffrey Knockel}
\orcid{https://orcid.org/0009-0006-6239-9725}
\affiliation{%
    \institution{Bowdoin College}
    \department{Department of Computer Science}
    \city{Brunswick}
    \state{ME}
    \postcode{04011}
    \country{USA}}
\email{j.knockel@bowdoin.edu}

\author{Jedidiah R. Crandall}
\orcid{https://orcid.org/0000-0001-7864-2992}
\affiliation{%
    \institution{Arizona State University}
    \department{Biodesign Center for Biocomputing, Security and Society}
    \department{School of Computing and Augmented Intelligence}
    \streetaddress{727 E. Tyler St.}
    \city{Tempe}
    \state{AZ}
    \postcode{85281}
    \country{USA}}
\email{jedimaestro@asu.edu}

\ccsdesc[500]{Security and privacy~Network security}
\ccsdesc[500]{Networks~Routing protocols}

\keywords{IP identifiers, side channels, comparative analysis, probability theory}

\acmJournal{CSUR}

\begin{abstract}
    The battle for a more secure Internet is waged on many fronts, including the most basic of networking protocols.
    Our focus is the \textit{IPv4 Identifier} (IPID), an IPv4 header field as old as the Internet with an equally long history as an exploited side channel for scanning network properties, inferring off-path connections, and poisoning DNS caches.
    This article taxonomizes the 25-year history of IPID-based exploits and the corresponding changes to IPID selection methods.
    By mathematically analyzing these methods' correctness and security and empirically evaluating their performance, we reveal recommendations for best practice as well as shortcomings of current operating system implementations, emphasizing the value of systematic evaluations in network security.
\end{abstract}

\begin{document}

\maketitle

\renewcommand{\shortauthors}{Daymude, Espinoza, Bergen, Mixon--Baca, Knockel, and Crandall}

\section{Introduction} \label{sec:intro}

There is ongoing interplay between operating system (OS) developers implementing basic networking protocols, such as IP and TCP, and security researchers discovering side channels and vulnerabilities in those diverse implementations.
A prime example is IP fragmentation and reassembly, a staple of Internet functionality for over forty years~\cite{RFC791}.
When an IPv4 packet is too large for the next link of its routing path, it is \textit{fragmented}; a destination machine then \textit{reassembles} fragments it receives based on their \textit{IPv4 Identifiers} (IPIDs), a 16-bit header field.
OSes are free to assign IPIDs to packets however they want, so long as they avoid causing ambiguous reassembly by sending multiple packets with the same IPID to the same destination close in time.

Despite early concerns about IP fragmentation's performance impacts~\cite{Kent1987-fragmentationconsidered} and later concerns about its vulnerabilities~\cite{Herzberg2013-fragmentationconsidered,Gilad2013-fragmentationconsidered}, its central role in the Internet---and in DNS in particular---continues to drive the discovery of new vulnerabilities in IPID selection methods.
Since 1998, numerous exploits have abused these methods to poison DNS caches~\cite{Herzberg2013-fragmentationconsidered,Palmer2019-firsttrydns,Zheng2020-poisontroubled}, hijack TCP connections~\cite{Feng2020-offpathtcp,Feng2022-offpathtcp,Gilad2012-offpathattacking}, launch denial-of-service (DoS) attacks~\cite{Malhotra2016-attackingnetwork}, scan ports~\cite{Antirez1998-newtcp,Zhang2018-onisinferring}, detect and measure connections~\cite{Knockel2014-countingpackets,Alexander2019-detectingtcp}, detect Internet censorship~\cite{Ensafi2014-detectingintentional}, and create covert channels~\cite{Klein2022-subvertingstateful}---all from off-path vantage points requiring nothing more than an active Internet connection.
Understanding the aims and necessary conditions of these diverse exploits is critical for proactively securing the menagerie of existing IPID selection methods from further misuse.

The purpose of this survey is two-fold.
First, we taxonomize the 25-year history of off-path IPID-based exploits and the subsequent changes to IPID selection methods across all major OSes.
This taxonomy has two levels, categorizing exploits primarily by their mechanisms (how does an exploit use IPIDs?) and secondarily by their end goals (what does the exploit do?).
Specifically, we demonstrate that despite the number and apparent diversity of IPID-based exploits, there are only two core mechanisms in use: (1) \textit{probe comparisons} that track changes in IPIDs over time to infer information about other machines, and (2) \textit{fragment injections} that use guessed IPIDs to trick a target machine into replacing legitimate fragments with malicious ones during reassembly.
For each mechanism, we identify the archetypal approach and resulting necessary conditions for success, thus distilling the key properties of (in)secure IPID selection methods.

Second, we perform a unifying comparative analysis of all seven major IPID selection methods across three qualities: \textit{correctness} (does the method supporting unambiguous reassembly by avoiding IPID collisions?), \textit{security} (are sequences of IPID values sufficiently difficult to predict?), and \textit{performance} (what are the method's time and space complexities?).
By parameterizing this analysis in terms of a machine's expected rate of traffic and number of CPUs, we reveal tradeoffs among the selection methods for different use cases (e.g., a high-traffic DNS server vs.\ a low-traffic home computer).
Among these evaluations, the most surprising is that globally incrementing selection---the first, simplest, and most dismissed of all IPID selection methods---is in fact the most collision-avoidant, secure, and performant choice for non-connection-bound packets when the rate of outgoing packets is high.
We conclude by proposing a new approach to IPID selection that shifts away from single-method implementations that necessarily compromise at least one of correctness, security, or performance across different use cases, instead embracing multiple implementations that end users can choose from (e.g., as a network setting) according to their specific needs.

The remainder of this paper is organized as follows.
Section~\ref{sec:background} introduces IP fragmentation and reassembly, the seven primary IPID selection methods, and their current OS implementations.
Section~\ref{sec:attacks} reviews the history of exploits using IPIDs for measurements or malicious attacks and the corresponding changes to selection methods.
Section~\ref{sec:comparison} presents our comparative analysis of IPID selection methods' correctness, security, and performance, summarized in Table~\ref{tab:comparison}.
We synthesize these evaluations as recommendations for a new approach to IPID selection in Section~\ref{sec:recommend}; readers seeking to quickly digest ``what's new'' should skip to this section and refer to \figtext~\ref{fig:phases} as a visual summary of our proposed approach.
Finally, we conclude in Section~\ref{sec:conclude}.

\section{Background} \label{sec:background}

We begin by reviewing how IPIDs are used, the established methods for selecting IPIDs, and the current OS implementations of those selection methods.

\subsection{IP Packet Fragmentation and Reassembly} \label{subsec:use}

\begin{figure}[t]
    \centering
    \includegraphics[width=0.7\textwidth]{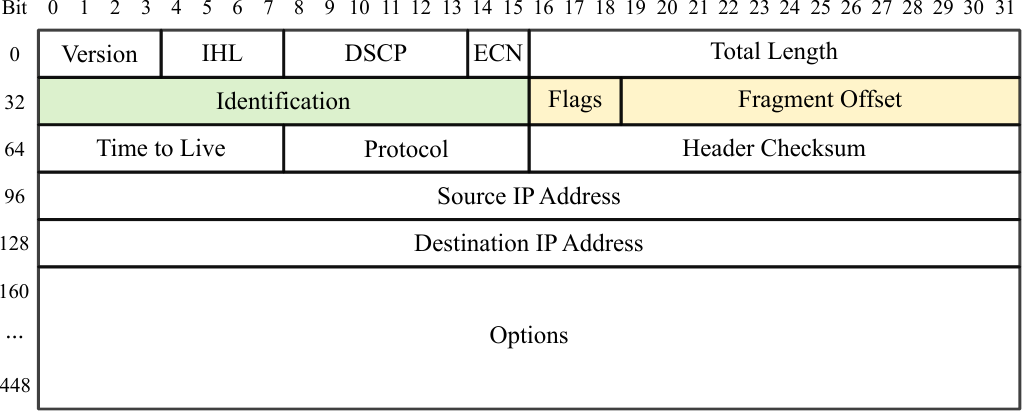}
    \caption{\textit{Anatomy of an IPv4 Packet Header.}
    The 16-bit Identification field is the packet's IPID.
    The Flags and Fragment Offset fields are also used in IP fragmentation and reassembly.}
    \label{fig:ipheader}
\end{figure}

The IPID is a 16-bit IPv4 header field used for packet reassembly (see \figtext~\ref{fig:ipheader}).
Every IPv4 packet is assigned an IPID by its sender.
If ever a packet is too large for some link of its routing path, an intermediate router breaks it into \textit{fragments} that inherit the IPID of their packet.
As the packet's destination receives these fragments in its \textit{reassembly buffer} (also known as the \textit{fragmentation cache}), it uses their IPIDs to group them by packet and inspects their \textit{Flags} and \textit{Fragment Offset} IPv4 header fields to determine whether all of the packet's fragments have arrived and in what order they should be reassembled.
This continues until the packet is reassembled or a timeout of 15--120 seconds is reached~\cite{RFC1122,RFC791}.
The main restriction concerning IPID selection is that IPIDs must be ``unique'' to avoid ambiguous reassembly~\cite{RFC791}.

IP fragmentation is not always necessary.
For example, if a sender knows a path's maximum transmission unit (MTU)---e.g., as part of a TCP connection---it can resize its packets prior to sending them.
Such packets can be made \textit{atomic} to discourage IP fragmentation, setting their \textit{Don't Fragment} (DF) flag to \true, their \textit{More Fragments} (MF) flag to \false, and their \textit{Fragment Offset} to 0.
IPIDs have no utility in atomic packets and are thus ignored and allowed to be any value~\cite{RFC6864}.

In IPv6, senders are ``strongly recommended'' to avoid fragmentation by either implementing path MTU discovery and resizing packets accordingly or limiting their packets to \numprint{1280} octets, the IPv6 minimum link MTU~\cite{RFC8200}.
When fragmentation does occur in IPv6, it is performed only by senders; intermediate routers never fragment IPv6 packets.
If a router determines that an IPv6 packet's size exceeds the next link's MTU, it drops the packet and sends an ICMPv6 ``Packet Too Big'' message back to the sender.
To send an IPv6 packet whose size exceeds the path MTU, a sender can split the packet into fragments and send each fragment as a separate IPv6 packet including the optional \textit{Fragment} extension header.
This extension header includes a 32-bit identification value used by the receiver during reassembly, analogous to the IPID in IPv4.
We primarily focus on IPv4's IPID in this survey, but will briefly return to IPv6 fragmentation in Section~\ref{subsec:ipv6attacks}.

\subsection{IPID Selection Methods} \label{subsec:select}

In this section, we detail the seven existing methods for selecting packets' IPIDs.
These methods fall into two categories: counter-based (Section~\ref{subsubsec:counter}) and PRNG-based (Section~\ref{subsubsec:prng}).

\subsubsection{Counter-Based Selection Methods} \label{subsubsec:counter}

\begin{itemize}
    \item \textit{Globally incrementing} IPID selection (Algorithm~\ref{alg:benchmark}, Lines~\ref{alg:benchmark:globalstart}--\ref{alg:benchmark:globalend}) is the earliest and simplest method.
    Each machine maintains a single, sequentially-incrementing, 16-bit counter for all IPID generation.
    Whenever a machine sends a packet, it increments its global counter (modulo $2^{16}$) and uses the resulting value as the packet's IPID.

    \item \textit{Per-connection} IPID selection (Algorithm~\ref{alg:benchmark}, Lines~\ref{alg:benchmark:connstart}--\ref{alg:benchmark:connend}) generalizes globally incrementing selection by maintaining one counter per active \textit{connection}, i.e., a 5-tuple specifying the source and destination IP addresses, source and destination ports, and protocol number.
    Each connection counter is initialized with a random 16-bit value when the connection is established and subsequently assigns IPIDs to packets by incrementing by one, modulo $2^{16}$.

    \item \textit{Per-destination} IPID selection (Algorithm~\ref{alg:benchmark}, Lines~\ref{alg:benchmark:deststart}--\ref{alg:benchmark:destend}) is analogous to per-connection selection, but with one counter per pair of source and destination IP addresses.

    \item \textit{Per-bucket} IPID selection (Algorithm~\ref{alg:benchmark}, Lines~\ref{alg:benchmark:bucketmutexstart}--\ref{alg:benchmark:bucketlinuxend}) is similar to per-connection and per-destination selection, but with one counter per ``bucket''.
    Packets are typically mapped to buckets by hashing their IP header information (e.g., source and destination IP addresses or the protocol number) with cryptographic information like secret keys.
    Some implementations use sequentially-incrementing bucket counters, analogous to the above methods, while others inject some stochasticity.
    These variations are detailed by OS in Section~\ref{subsec:implement}.
\end{itemize}

\subsubsection{PRNG-Based Selection Methods} \label{subsubsec:prng}

Methods based on \textit{pseudorandom number generation} (PRNG) select IPIDs at random from among the $2^{16}$ possible values, optionally employing some mechanism for avoiding recently-selected IPIDs whose reuse could cause ambiguous reassembly.
Naturally, these methods' quality depend on unbiased, secure PRNG algorithms drawing from good sources of entropy.
Cryptanalysis of PRNGs (e.g.,~\cite{Gutterman2006-analysislinux,Dorrendorf2009-cryptanalysisrandom}) is beyond the scope of this survey, but we will discuss examples of poor-quality PRNGs affecting IPID selection in Sections~\ref{subsubsec:idlescans} and~\ref{subsubsec:dnsattacks}.

\begin{itemize}
    \item One method for generating non-repeating random IPIDs is with a \textit{searchable queue} storing the last $k \geq 0$ unique IPIDs (Algorithm~\ref{alg:benchmark}, Lines~\ref{alg:benchmark:prngqueuestart}--\ref{alg:benchmark:prngqueueend}).
    When an IPID is requested, the PRNG generates 16-bit values until one is found that is not already in the queue; this new value is then inserted into the queue and used as the next IPID.
    The size of the searchable queue varies by implementation, but is typically chosen between \numprint{4096} and \numprint{32768} (i.e., $2^{12}$ and $2^{15}$) to provide a good tradeoff between entropy and non-repetition.

    \item Another method uses an \textit{iterated Knuth shuffle}~\cite{Knuth1969-artcomputer}, also known as the Durstenfeld algorithm for random permutations or the Fisher--Yates shuffle (Algorithm~\ref{alg:benchmark}, Lines~\ref{alg:benchmark:prngknuthstart}--\ref{alg:benchmark:prngknuthend}).
    Initially, a random permutation of all $2^{16}$ IPID values is generated by a standard Knuth shuffle.
    An index $i$ marks the current ``head'' of the (cyclic) permutation.
    When an IPID is requested, $i$ is incremented and the $i$-th value in the permutation is used.
    That value is then swapped into a position chosen uniformly at random from $[i - (2^{16} - k - 1) \bmod 2^{16}, i]$ (i.e., among the previous $2^{16} - k$ positions in the cyclic permutation, including its own), ensuring that it will not be selected again within the next $k$ requests for IPIDs.

    \item \textit{Pure PRNG selection} simply generates IPIDs from among the $2^{16}$ possible values uniformly at random, possibly repeating IPIDs.
    This is functionally equivalent to the previous two methods with $k = 0$, but has no need for their data structures ensuring non-repetition.
\end{itemize}

\subsection{Current Implementations} \label{subsec:implement}

In this section, we survey current implementations of IPID selection across popular OSes.

\paragraph{Windows}

Source code for Windows 8, 10, and 11 is not publicly available, though reverse-engineering efforts revealed that until related issues were patched in 2019, Windows 8 and 10 used per-bucket IPID selection based on the Toeplitz hash function with \numprint{8192} (i.e., $2^{13}$) buckets~\cite{Klein2019-ipid}.
Recent analysis shows that Windows Server (v1903) uses per-destination IPID selection~\cite{Klein2022-subvertingstateful}.
Specifically, Windows maintains a ``PathSet'' (i.e., a hash table) of sequentially incrementing counters indexed by source and destination IP address pairs, adding a new counter whenever a packet's destination is not already in the PathSet.
The PathSet's size is checked every 0.5~s: if it has grown beyond its ``purge threshold''---\numprint{4096} (i.e., $2^{12}$) counters in Windows 10 or \numprint{32768} (i.e., $2^{15}$) in Windows Server---or more than \numprint{5000} counters have been added since the last check, a ``purge sequence'' removes up to $\max\{\numprint{1000}, \text{\# added since last check}\}$ ``stale'' counters.
Counters are considered ``stale'' depending on when they were last accessed and the PathSet's size: if the PathSet's size is between once and twice its purge threshold, then counters accessed longer than 10~s (Windows 10) or 60~s (Windows Server) ago are considered stale; if the PathSet's size is more than twice its purge threshold, all counters are considered stale~\cite{Klein2024-privatecommunication}.

\paragraph{Linux}

The Linux kernel (\href{https://elixir.bootlin.com/linux/v6.9/source/include/net/ip.h#L542}{v6.9}) uses either per-connection or per-bucket IPID selection depending on the connection type.
Packets sent via established sockets use per-connection selection, initializing connection counters with a random value generated by \texttt{prandom\_u32} after the connection is established.
Atomic packets sent outside of sockets and any RST response to an unsolicited packet (e.g., a stray SYN/ACK) are assigned IPID zero~\cite{Alexander2019-detectingtcp}.
All other packets use per-bucket selection with \numprint{2048} to \numprint{262144} (i.e., $2^{11}$ to $2^{18}$) buckets, depending on the sender's RAM.
A packet is hashed to a bucket using SipHash-2-4 on its destination IP address, source IP address, protocol number, and a 128-bit random key.
Instead of incrementing bucket counters sequentially (i.e., by one, modulo $2^{16}$), Linux uses \textit{stochastic increments} based on the number of system ticks since a bucket counter was last incremented.
Formally, if a bucket counter with value $c$ was last accessed at time $t_\text{old}$ and is being accessed for a new IPID at time $t_\text{now}$ (where times are expressed in system ticks), then an increment $inc$ is sampled uniformly at random from $[1, \max\{1, t_\text{now} - t_\text{old}\}]$ and the new IPID and bucket counter value is $c + inc \bmod 2^{16}$.
These stochastic increments were designed to add noise to less frequently used bucket counters that adversaries could otherwise take advantage of.

\paragraph{OpenBSD}

OpenBSD (\href{https://github.com/openbsd/src/blob/085f25454964baa8e68d1053b787c22afc566ce9/sys/netinet/ip_id.c}{v7.5}) implements PRNG IPID selection using an iterated Knuth shuffle with $k = \numprint{32768}$ (i.e., $2^{15}$).
Random swap indices are generated using \texttt{arc4rand}.
OpenBSD never assigns zero as an IPID: if zero is next in the permutation, it is swapped as usual but not returned.

\paragraph{FreeBSD}

FreeBSD (\href{https://github.com/freebsd/freebsd-src/blob/release/14.0.0/sys/netinet/ip_id.c}{v14.0}) assigns IPID zero to all atomic packets and uses globally incrementing selection for non-atomic packets.
FreeBSD includes a configuration option to use PRNG selection with a searchable queue of size $k = \numprint{8192}$ (i.e., $2^{13}$) instead, though it is disabled by default to minimize performance impact.
This disabled code uses \texttt{arc4rand} as its PRNG.
As in OpenBSD, zero is treated as a special IPID that is never returned.

\paragraph{macOS/XNU}

Although most source code for macOS is not publicly available, its network stack uses the open source XNU kernel (\href{https://github.com/apple-oss-distributions/xnu/blob/xnu-8792.41.9/bsd/netinet/ip_id.c}{v8792.41.9}).
As in FreeBSD, macOS/XNU assigns zero as the IPID of all atomic packets.
For non-atomic packets, macOS recently adopted pure PRNG selection, generating 16-bit values uniformly at random and then salting them with a packet-specific value; notably, this implementation does not include mechanisms for avoiding repeated IPIDs.
As in the BSDs, zero is treated as a special IPID that is never returned.

\section{Measurements and Attacks} \label{sec:attacks}

\begin{figure}[t]
    \centering
    \includegraphics[width=\textwidth]{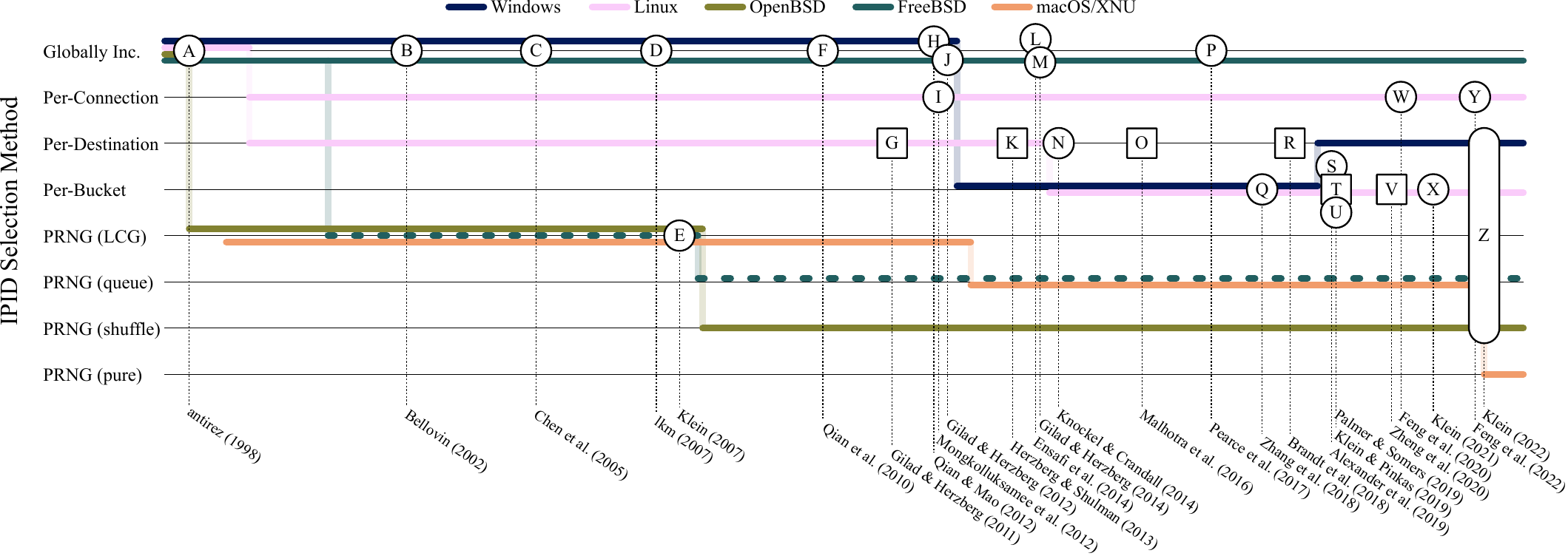}
    \vspace{-7.5mm}
    \caption{\textit{Timeline of IPID Measurements, Attacks, and Selection Methods.}
    Each OS implementation of IPID selection varies over time in response to disclosed measurements and attacks (Section~\ref{sec:attacks}), categorized either as probe comparisons (circles) or fragment injections (squares).
    Since 2000, Linux has implemented two methods, one for atomic packets (per-connection) and one for non-atomic packets (per-destination or per-bucket).
    FreeBSD has implemented various PRNG-based methods (dashed lines) but always disables them by default.
    Minor patches that did not change selection methods (e.g., a different hash function in per-bucket selection or Linux's fixes for specific packet types) are not shown.}
    \label{fig:timeline}
\end{figure}

In this section, we present a history of IPID-based exploits and the corresponding OS implementation changes they influenced (see \figtext~\ref{fig:timeline} for an overview).
Notably, many vulnerabilities remained exploitable long after disclosure due to delays in patching and updating to the latest versions.
We categorize these exploits as either \textit{probe comparisons} (Section~\ref{subsec:probecompare}) or \textit{fragment injections} (Section~\ref{subsec:fraginject}) based on how they use IPIDs to achieve their goals.
Within these categories, we distinguish between \textit{measurements} which infer information about other machines and \textit{attacks} which maliciously affect target systems---often by leveraging information learned from measurements.
We conclude with a brief discussion of related exploits against IPv6 fragmentation (Section~\ref{subsec:ipv6attacks}).

Network exploits by a machine $M$ against target machines $A$ and $B$ are performed from one of three vantage points: off-path, on-path, or in-path~\cite{Marczak2015-analysischina}.
If $M$ is \textit{off-path}, $M$ does not receive any traffic between $A$ and $B$.
Instead, $M$ must actively \textit{probe} these machines by sending them packets and inspecting their responses to learn anything about their states.
An \textit{on-path} machine passively receives all traffic between $A$ and $B$ but cannot modify it directly; an \textit{in-path} machine (i.e., a ``man-in-the-middle'') receives and can modify all traffic between $A$ and $B$.
For example, if an ISP were to forward copies of its traffic to its government, the ISP is in-path and the government is on-path.
We are primarily concerned with off-path exploits, since these rely on side channels like IPIDs for information.
They are also the most powerful in practice, since they are usable by any unprivileged adversary without ISP or nation-state control over network infrastructure.

\subsection{Probe Comparisons} \label{subsec:probecompare}

A \textit{probe comparison} tracks changes in IPIDs over time to infer information about other machines.
As such, most probe comparisons are measurements.
The inferred information is typically binary: whether a target port is open or closed, a firewall is performing egress or ingress filtering, a guessed secret value is correct or incorrect, and so on.
Probe comparisons come in many forms, but the archetypal approach for an off-path measurer or attacker works as follows:
\begin{enumerate}
    \item Probe the target machine's relevant IPID, obtaining value $x$.

    \item Interact with the target machine such that its IPID values change one way if the inferred information is ``true'' (e.g., the target port is open, a firewall is performing egress filtering, a guessed secret value is correct, etc.) and some other way otherwise.

    \item Probe the target machine's relevant IPID again, obtaining a new value $y$.

    \item Based on the interactions in Step (2), compare $x$ and $y$ to gather evidence about whether the inferred information is ``true''.
    Repeat Steps (1)--(3) as needed to gain statistical confidence.
\end{enumerate}
From this outline, we can characterize two necessary conditions for successful probe comparisons:
\begin{enumerate}[label=(PC\arabic*), ref=PC\arabic*]
    \item \label{cnd:pc:access} A measurer or attacker must be able to probe and interact with a target machine's relevant IPID selection mechanism, either directly or indirectly, in a timely manner.

    \item \label{cnd:pc:correlate} A measurer or attacker must be able to correlate its interactions with the target machine to observable changes in the target machine's IPID values.
\end{enumerate}

The remainder of this section summarizes seven probe comparison exploits grouped by their end goals, including NAT measurements (Section~\ref{subsubsec:onpathNAT}), idle scans (Section~\ref{subsubsec:idlescans}), triangular spamming (Section~\ref{subsubsec:trispam}), communication measurements (Section~\ref{subsubsec:canaryfragments}), bucket leaks (Section~\ref{subsubsec:bucketleaks}), TCP injection attacks (Section~\ref{subsubsec:tcpinject}), and covert channels (Section~\ref{subsubsec:covertchannels}).

\subsubsection{On-Path NAT Measurements} \label{subsubsec:onpathNAT}

It was once assumed that \textit{network address translation} (NAT)~\cite{RFC2663} obfuscated devices in a local network behind a shared public IP address, stopping external observers on the public network from counting or identifying NATted devices or distinguishing their traffic.
However, early NAT devices did not change outgoing packets' IPIDs.
Bellovin~\cite{Bellovin2002-techniquecounting} (\figtext~\ref{fig:timeline}B) showed how this could be exploited for on-path measurement (e.g., by an ISP that sees all NATted traffic but does not manipulate it) if consecutive IPIDs assigned by the same NATted device are correlated.
Notably, this on-path vantage point eliminates the need for probe comparison's probing and interaction steps, simplifying it to comparing IPIDs over time.
\ifnonacm
First, the on-path measurer collects all traffic from a target NAT over a period of time.
It then builds sequences of IPIDs that likely originated from the same NATted device by inspecting packets' IPIDs in time order and either appending them to the ``best match'' existing sequence or starting a new sequence if none match.
The resulting number of sequences is often a good approximation of the number of NATted devices.
\else\fi

Bellovin's
\ifnonacm
matching rules for constructing IPID sequences
\else
rules for matching IPID sequences to NATted devices
\fi
are based on globally incrementing IPIDs, but could easily extend to any counter-based method by changing parameters.
Thus, the paper's suggested mitigations included using a PRNG-based selection method.
However, this protection alone proved insufficient: Mongkolluksamee et al.~\cite{Mongkolluksamee2012-countingnatted} (\figtext~\ref{fig:timeline}I) later extended this on-path measurement to uniquely identify NATted devices and their OSes across all IPID selection methods by additionally observing patterns in the TCP sequence number and source port.

\subsubsection{Idle Scans} \label{subsubsec:idlescans}

\begin{figure}[t]
    \centering
    \begin{subfigure}{0.4\textwidth}
        \centering
        \includegraphics[width=\textwidth]{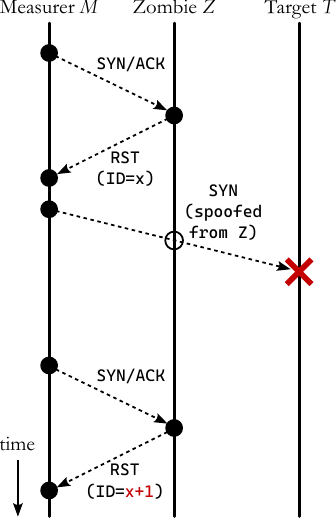}
        \caption{The targeted port on $T$ is closed.}
        \label{fig:idlescan:closed}
    \end{subfigure}
    \hspace{0.075\textwidth}
    \begin{subfigure}{0.4\textwidth}
        \centering
        \includegraphics[width=\textwidth]{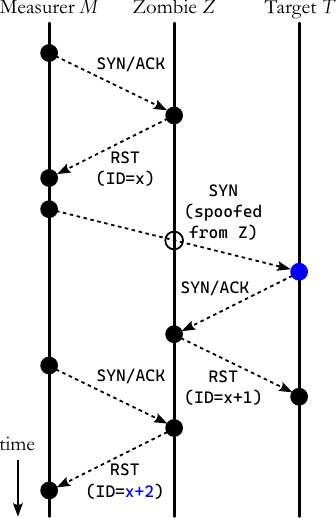}
        \caption{The targeted port on $T$ is open.}
        \label{fig:idlescan:open}
    \end{subfigure}
    \caption{\textit{Idle Scan Execution.}
    In an idle scan, a measurer $M$ is testing whether a target port on another machine $T$ is open.
    Measurer $M$ first probes the zombie machine $Z$'s current IPID, $x$.
    It then spoofs a SYN packet to the target machine $T$ that appears to have come from $Z$.
    (a) If the target port on $T$ is closed (red X), then $T$ does not respond and $M$'s final probe of $Z$'s IPID is $x+1 \bmod 2^{16}$.
    (b) Otherwise, if the target port on $T$ is open (blue circle), the resulting interaction between $T$ and $Z$ causes $M$'s final probe of $Z$'s IPID to be $x+2 \bmod 2^{16}$.}
    \label{fig:idlescan}
\end{figure}

In the previous section, an on-path vantage point enabled IPID-based measurements that could be performed entirely passively.
However, off-path measurements require the active transmission of packets, not just to probe current IPID values but often also to record the effect (or lack thereof) of a packet transmission on subsequent IPIDs.
The earliest known example of such a measurement is the \textit{idle scan} which exploits the predictability of globally incrementing IPIDs to reveal off-path connections~\cite{Antirez1998-newtcp,DeVivo1999-reviewport,Handley2001-networkintrusion} (\figtext~\ref{fig:timeline}A).
An off-path measurer $M$ can learn whether a target machine $T$ has a certain port open by probing the IPIDs of an intermediate ``zombie'' machine $Z$ with low traffic rates (\figtext~\ref{fig:idlescan}).
First, $M$ sends $Z$ an unsolicited SYN/ACK packet, to which $Z$ replies with an RST packet that contains its current IPID, say $x$.
Machine $M$ then sends a spoofed SYN packet to $T$ appearing to originate from $Z$ destined for the target port it wants to scan.
If this port is closed, nothing comes of this spoofed packet (\figtext~\ref{fig:idlescan:closed}); otherwise, if it is open, $T$ replies to $Z$ with a SYN/ACK, which in turn causes $Z$ to reply to $T$ with an RST containing an incremented IPID (\figtext~\ref{fig:idlescan:open}).
Finally, $M$ again sends $Z$ an unsolicited SYN/ACK and inspects the incremented IPID $y$ of the RST reply.
If $y \equiv x + 2 \bmod 2^{16}$, $M$ concludes the target port on $T$ is open; otherwise, if $y \equiv x + 1 \bmod 2^{16}$, it concludes the port is closed or filtered.

As a result of the idle scan's disclosure, OpenBSD, FreeBSD, and XNU replaced their globally incrementing IPIDs with PRNG selection using Linear Congruential Generators (LCGs) with additional complication layers.
These implementations lasted until 2007 when Klein~\cite{Klein2007-openbsddns} (\figtext~\ref{fig:timeline}E) demonstrated that the ``random'' values produced by LCGs could be predicted, once again making idle scans possible.
In response, OpenBSD replaced LCGs with an iterated Knuth shuffle (see OpenBSD commit \href{https://github.com/openbsd/src/commit/4fd19fd8c5fe0efc775193f1f0ab218383fe8fcd}{4fd19f}) while FreeBSD---and later, XNU---chose PRNG selection using searchable queues (see FreeBSD commit \href{https://github.com/freebsd/freebsd-src/commit/361021cc6ee359629b21df3e29c14544d05a38ff}{361021} and XNU commit \href{https://github.com/apple/darwin-xnu/commit/ff3a0c1e9808972592661c365ab368466b31ab20}{ff3a0c}).
These implementations remain largely unchanged since their appearances in 2008.
Linux took a different approach, implementing a hybrid of per-connection selection (for atomic packets) and per-destination selection (for non-atomic packets).
Windows retained globally incrementing IPIDs until Windows 8, which was released in 2012~\cite{Orevi2018-dnsdnsdnsbased}.

Despite these changes, delayed patching gave ample opportunity for the idle scan's wide adoption and refinement.
Nmap integrated it as a scanning option in 2001~\cite{Lyon2009-nmapnetwork}.
Chen et al.~\cite{Chen2005-exploitingipid} (\figtext~\ref{fig:timeline}C) introduced an extension that additionally inferred packet loss, duplication, and arrival order; server traffic rates; and the number of servers in a load-balanced deployment.
Later, Ensafi et al.~\cite{Ensafi2014-detectingintentional} (\figtext~\ref{fig:timeline}L) improved the idle scan to detect off-path SYN/ACK and RST filtering between machines~\cite{Ensafi2010-idleport}, which in turn was leveraged for Internet-wide censorship measurement~\cite{Ensafi2014-advancednetwork,Ensafi2015-analyzinggreat,Pearce2017-augurinternetwide} (\figtext~\ref{fig:timeline}P).

\subsubsection{Triangular Spamming} \label{subsubsec:trispam}

Qian et al.~\cite{Qian2010-investigationtriangular} (\figtext~\ref{fig:timeline}F) formalized \textit{triangular spamming}, a technique for email spam where a high-bandwidth spammer spoofs spam through low-bandwidth relay bots to a target mail server, thus concealing its own IP address and protecting itself from IP block-listing.
IPID probe comparisons are less relevant to the actual execution of triangular spamming, but play a key role in its setup when the attacker identifies susceptible ISPs and relay bots.
Specifically, the relay bot(s) must be able to receive responses inbound on port 25 from the mail server and forward them to the spammer using some other port; i.e., the ISP must not be ingress filtering on port 25.

If a potential relay bot $R$ uses globally incrementing IPIDs, the presence of ingress filtering on port 25 can be determined using a simple probe comparison.
The attacker $A$ first probes $R$ with an unsolicited SYN/ACK on port 80 and records the IPID $x$ of the RST reply.
It then sends a large burst of $s \gg 1$ probes on port 25 which each increment the global IPID counter on $R$ if and only if inbound traffic on port 25 is not blocked.
Finally, $A$ again probes $R$ on port 80 to obtain its updated IPID $y$.
If $y \approx x + 1 \bmod 2^{16}$, the ISP is likely blocking inbound traffic; otherwise, if $y \approx x + s \bmod 2^{16}$, the relay bot is receiving packets on port 25 and can be used in triangular spamming.

\subsubsection{Off-Path Measurement of Inter-Machine Communication} \label{subsubsec:canaryfragments}

\begin{figure}[t]
    \centering
    \includegraphics[width=0.6\textwidth]{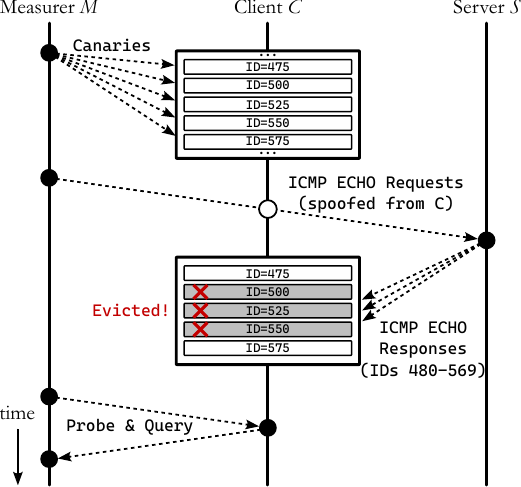}
    \caption{\textit{Canary Fragments Execution.}
    If a server $S$ uses a sequentially incrementing counter for its IPIDs, a measurer $M$ can count packets sent from $S$ to a client $C$ by planting canary fragments with a range of guessed IPIDs in the reassembly buffer of $C$ and later querying whether they were evicted.}
    \label{fig:canaryfragments}
\end{figure}

For 14 years, Linux's per-destination IPIDs for non-atomic packets seemed impervious to probe comparisons.
If an off-path measurer $M$ probes the per-destination counter on a server $S$ for some client $C$ by spoofing a packet from $C$ to $S$, the server $S$ replies to $C$, not $M$.
But this intuition also reveals a weakness: every increment of a per-destination counter is caused by a packet sent to its unique destination.
Knockel and Crandall~\cite{Knockel2014-countingpackets} (\figtext~\ref{fig:timeline}N) recognized that this could be abused to detect if $S$ and $C$ are communicating and, if so, count the number of ICMP, UDP, or TCP packets exchanged between them.
First, $M$ fills the reassembly buffer of $C$ with \textit{canary fragments} containing guessed IPIDs that appear to originate from $S$ (\figtext~\ref{fig:canaryfragments}).
Second, $M$ spoofs large ICMP ECHO requests from $C$ to $S$ which become fragmented ICMP ECHO responses from $S$ to $C$.
If any canary has an IPID matching an ECHO response, it is evicted from the reassembly buffer.
Finally, $M$ ``queries'' the reassembly buffer of $C$ to determine if many canaries are missing; if so, their guessed IPIDs were correct.
By repeating this measurement over time and tracking the change in IPIDs, $M$ can count the number of packets sent from $S$ to $C$.

When canary fragments were disclosed, the Linux developers had already been working to replace per-destination selection---due to its large performance impact on modern servers---with per-bucket selection using a Jenkins hash of the destination IP address (see Linux commit \href{https://github.com/torvalds/linux/commit/73f156a6e8c1074ac6327e0abd1169e95eb66463}{73f156}).
To curb canary fragments' abuse of per-destination counters' perfect increments-to-packets correlation, they also added the stochastic increments described in Section~\ref{subsec:implement} and included the source IP address and protocol number in the bucket hash input (see Linux commit \href{https://github.com/torvalds/linux/commit/04ca6973f7c1a0d8537f2d9906a0cf8e69886d75}{04ca69}).

\subsubsection{Bucket Leaks} \label{subsubsec:bucketleaks}

Conceptually, per-bucket selection blends globally incrementing selection's noisiness of shared counters with per-destination selection's difficulty of direct probing, but in practice has proven vulnerable to the side channels of both.
Recall that a bucket hash function $h$ hashes packets to buckets using a packet's source and destination IP addresses $s$ and $d$, the protocol number $p$, and a secret key $k$.
If a measurer controls an IP address $d'$ such that $h(s, d, p, k) = h(s, d', p, k)$, it can exploit this \textit{hash collision} to probe the per-bucket counter assigning IPIDs to machine $d$.
Below, we summarize three measurements exploiting this side channel.

\paragraph{Non-Idle Port Scans}

Zhang et al.~\cite{Zhang2018-onisinferring} (\figtext~\ref{fig:timeline}Q) introduced the first method for discovering bucket hash collisions and used it to perform off-path port scans without an idle machine.
To find a hash collision, a measurer $M$ first sends an unsolicited SYN/ACK to a server $S$ and records the RST reply's IPID, say $x$, which is drawn from bucket $b_M = h(S, M, p, k)$.
Measurer $M$ then spoofs a SYN/ACK to $S$ that appears to originate from the client $C$ that $M$ is trying to collide with.
Server $S$ replies to $C$ with an RST whose IPID is drawn from bucket $b_C = h(S, C, p, k)$.
Finally, $M$ again probes bucket $b_M$ of $S$ to obtain IPID $y$.
Importantly, $M$ sends all three SYN/ACKs in rapid succession to avoid the per-bucket counter's stochastic increments.
If $y \equiv x+2 \bmod 2^{16}$, $M$ concludes that $b_M = b_C$; otherwise, if $y \equiv x+1 \bmod 2^{16}$, $M$ and $C$ are likely hashing into different buckets.

On failing to find a hash collision, $M$ can repeat this process with a different IP address it controls.
The probability of finding a hash collision for one of $B$ buckets when controlling $D$ IP addresses is $1 - (1 - 1/B)^D$, which is over $63\%$ when $D \approx B$ for the typical range of $B \in [2^{11}, 2^{18}]$.
At the time this scan was discovered, similar collision probabilities were achievable with fewer IPv4 addresses when targeting dual-stack machines (those with both IPv4 and IPv6 addresses), since it is relatively easy to obtain a large block of IPv6 addresses and the same counters were used for both per-bucket IPIDs and IPv6 fragment identifiers.
Now, Linux generates purely random, non-zero IPv6 fragment IDs (see Linux commit \href{https://github.com/torvalds/linux/commit/62f20e068ccc50d6ab66fdb72ba90da2b9418c99}{62f20e}), mitigating this dual-stack weakness and related issues~\cite{Klein2022-subvertingstateful}.

Once a collision is found, the remainder of the port scan operates like an idle scan (Section~\ref{subsubsec:idlescans}) with added statistical inference to overcome per-bucket counters' stochastic increments.
\ifnonacm
Measurer $M$ first probes the server $S$ with the typical SYN/ACK to obtain the per-bucket IPID $x_1$ at time $t_1$.
It then spoofs a SYN packet to the client $C$ from $S$.
As in the idle scan, this has no effect if the target port on $C$ is closed; otherwise, if it is open, $C$ responds to $S$ with a SYN/ACK, which in turn causes $S$ to respond to $C$ at time $t_2$ with an RST containing IPID $x_2 = x_1 + \mathcal{U}_{[1, t_2 - t_1]}$.
Finally, $M$ probes $S$ again to obtain IPID $x_3$ at time $t_3$.
If the target port on $C$ is closed, then $x_3 - x_1 = \mathcal{U}_{[1, t_3 - t_1]}$; otherwise, the additional IPID $x_2$ generated when $C$ and $S$ completed the SYN handshake means $x_3 - x_1 = \mathcal{U}_{[1, t_2 - t_1]} + \mathcal{U}_{[1, t_3 - t_2]}$.
These two distributions are distinct, so the status of the target port can be inferred by repeating measurements of $x_1$ and $x_3$ and applying an appropriate statistical test.
\else\fi

\paragraph{Hybrid Leaks}

Alexander et al.~\cite{Alexander2019-detectingtcp} (\figtext~\ref{fig:timeline}S) introduced \textit{hybrid leaks}, a measurement that detects active TCP connections between a Linux server $S$ and any of its clients $C$ by differentiating between Linux's hybrid modes of per-bucket and per-connection selection.
First, a measurer $M$ finds a hash collision with $C$ on $S$ using a simplified version of the Zhang et al.~\cite{Zhang2018-onisinferring} method described above.
It then sends an unsolicited SYN/ACK to $S$ and obtains the current per-bucket IPID $x$ from the RST reply.
Next, $M$ spoofs $s > 1$ SYN/ACKs from $C$ to $S$ containing a guessed source port and a standard TCP destination port (i.e., 80 or 443).
If $S$ and $C$ have an active TCP connection on those ports, $S$ replies to $C$ with a single ``challenge ACK'' whose IPID is drawn from their per-connection counter.
Otherwise, $S$ replies to $C$ with $s$ RST packets whose IPIDs each increment the per-bucket counter.
Finally, $M$ probes the updated per-bucket IPID $y$.
Assuming $M$ sends its packets fast enough to avoid the per-bucket counter's stochastic increments, $y - x \bmod 2^{16} < s$ is evidence that $S$ and $C$ have an active connection.

Linux mitigated this probe comparison measurement---and others depending on unsolicited SYN/ACK probes and their RST replies~\cite{Antirez1998-newtcp,Lkm2007-remoteblind,Qian2010-investigationtriangular,Zhang2018-onisinferring}---by replying to any unsolicited SYN/ACK with an RST whose IPID is zero (see Linux commit \href{https://github.com/torvalds/linux/commit/431280eebed9f5079553daf003011097763e71fd}{431280}).

\paragraph{Device Fingerprinting}

Klein and Pinkas~\cite{Klein2019-ipid} (\figtext~\ref{fig:timeline}U) leveraged a combination of reverse-engineering, bucket hash collisions, and cryptanalysis using sequences of IPID values to exfiltrate the secret keys Windows and Linux devices used in their bucket hash functions.
Since these secret keys are generated randomly at device startup and remain fixed until a device restart, they form a long-lived device fingerprint, enabling a measurer to track devices across browsers, private sessions, VPNs, and network configuration changes.
At a high level, the Windows exfiltration technique takes advantage of two facts: (1) the hash function Windows used at the time assigns all destination IP addresses from the same class B network to the same bucket counter, and (2) the same secret keys and a related hash function are used to calculate offsets that are added to a counter values before returning them as IPIDs.
This enables a measurer to probe a target bucket counter from a small number of IP addresses across carefully chosen class B networks and then substitute the returned IPIDs into a system of linear equations whose (very small) set of solutions necessarily contains a sufficiently large portion of the device's secret key to be used as a fingerprint.
The Linux exfiltration technique leverages bucket collisions in a different way, probing a target machine in bursts from a range of IP addresses and then collecting pairs of IPIDs that are probabilistically likely to have originated from the same bucket.
All possible secret keys are then exhaustively searched, and the one predicting the highest number of collision pairs is considered the true key.

In response to this exploit, Linux developers increased the secret key size from 64 to 128 bits and updated the hash function to SipHash-2-4, which remains the current implementation (see Linux commit \href{https://github.com/torvalds/linux/commit/df453700e8d81b1bdafdf684365ee2b9431fb702}{df4537}).
Notably, Windows chose to address this issue by replacing per-bucket selection with per-destination~\cite{Klein2022-subvertingstateful} which, as already discussed in Section~\ref{subsubsec:canaryfragments}, has a high performance impact and has been exploited successfully before~\cite{Knockel2014-countingpackets,Malhotra2016-attackingnetwork}.

\subsubsection{TCP Injection Attacks} \label{subsubsec:tcpinject}

Beyond measuring information that an off-path observer would not otherwise be privy to, probe comparisons can also lay the groundwork for attacks that control or affect resources typically assumed to be beyond an off-path attacker's influence.
In a \textit{TCP injection attack}, an attacker inserts malicious traffic into an active TCP connection between a client and server.
To do so, the attacker needs to know the client and server's IP addresses and TCP ports along with the current TCP sequence number of the machine the attacker is spoofing.
Most attacks assume the two IP addresses are known and the server is using a standard port for TCP (e.g., 80 or 443), leaving the client's port and sequence number as the two unknowns.
But if the client uses globally incrementing IPIDs, an off-path attacker can learn the client's port and sequence number through a port comparison guess-and-check~\cite{Lkm2007-remoteblind} (\figtext~\ref{fig:timeline}D).
The attacker (1) probes the client's IPID, say $x$, (2) sends packets that are specifically crafted to make the client increment its IPID if and only if its guessed port or sequence number is correct, (3) probes the client's new IPID, say $y$, and (4) compares $x$ and $y$ to determine the accuracy of its guess.
Just like for idle scans (Section~\ref{subsubsec:idlescans}), this attack's second step depends critically on the connection having little traffic, since other packets incrementing the client's IPID would confuse the results.
However, Gilad and Herzberg~\cite{Gilad2012-offpathattacking,Gilad2014-offpathtcp} (\figtext~\ref{fig:timeline}J,M) showed that with a ``puppet'' on the client (e.g., a malicious script in a browser sandbox), this attack can succeed even on noisier connections.

Naturally, the defense against this guess-and-check technique and the resulting TCP injections is to use an IPID selection method that is less predictable or harder for an attacker to access.
However, this must be done on all machines involved in the connection, not just the client.
Qian and Mao~\cite{Qian2012-offpathtcp} (\figtext~\ref{fig:timeline}H) showed that even if a client's IPIDs are difficult to predict, a firewall or middlebox using globally incrementing IPIDs can leak the sequence numbers necessary for TCP injection.

The state-of-the-art attacks remained specific to globally incrementing IPIDs for 13 years, until Feng et al.~\cite{Feng2020-offpathtcp,Feng2022-offpathtcp,Feng2022-pmtudnot} (\figtext~\ref{fig:timeline}W,Y) introduced their novel \textit{downgrade attacks} to achieve TCP injection against Linux's hybrid per-connection/per-bucket IPIDs.
Recall that when guess-and-checking the client's TCP port and sequence numbers, the attacker needs to probe the client's relevant IPIDs.
But it is impossible for the attacker to probe the per-connection counter specific to this client--server's active TCP connection unless it has the very information it's trying to leak.
Instead, Feng et al.\ observe that per-connection selection can be downgraded to per-bucket selection by spoofing ICMP ``Fragmentation Needed'' packets from the client to the server, tricking it into clearing its responses' DF flags and thus demoting otherwise atomic packets to non-atomic.
The client's corresponding bucket counter can then be identified and probed using the hash collision techniques described in Section~\ref{subsubsec:bucketleaks}, and an appropriately modified version of the above guess-and-check technique will leak the client port and sequence number(s) needed for TCP injection.
Feng et al.\ demonstrated that this method can be used to poison HTTP and BGP traffic in the wild~\cite{Feng2022-pmtudnot}.
Linux mitigated this issue in a case-specific manner, replacing per-bucket selection with an IPID generated by \texttt{prandom\_u32} only for SYN/ACKs that are large enough to be fragmented (see Linux commit \href{https://github.com/torvalds/linux/commit/970a5a3ea86da637471d3cd04d513a0755aba4bf}{970a5a}).

\subsubsection{Covert Channels} \label{subsubsec:covertchannels}

A \textit{covert channel} is an attack that transfers data between two processes or machines that are not supposed to be able to communicate, e.g., a privileged program leaking data to an unprivileged user~\cite{Lampson1973-noteconfinement}.
Klein~\cite{Klein2022-subvertingstateful} (\figtext~\ref{fig:timeline}Z) demonstrated that both counter-based and PRNG-based IPIDs can be used as covert channels to transmit information from a sender within an isolated network to any external receiver.
Although the details differ across OSes and IPID selection methods, the main idea is for the isolated sender to force its outside-facing firewall or host machine to either send a number of packets (signalling a 1 bit) or not (signalling a 0 bit), which the external receiver can view via probe comparison of the firewall/host's IPIDs.

Perhaps the most surprising of these covert channels uses PRNG selection with a searchable queue of size $k$, since PRNG IPIDs' unpredictability is not a defense in this case.
First, the receiver quickly sends $k$ probes to the firewall/host so that the replies' IPIDs $x_1, \ldots, x_k$ are exactly those in the searchable queue.
The isolated sender then communicates a 0 or 1 bit as described above.
Finally, the receiver sends another $k$ probes to the firewall/host to obtain IPIDs $y_1, \ldots, y_k$.
If there is overlap between the $x_i$'s and $y_i$'s, some packets must have been sent between the two sets of probes to evict the original members in the queue, likely indicating that the sender communicated a 1 bit; otherwise, if the $x_i$'s and $y_i$'s are disjoint, the sender likely communicated a 0 bit.

All major OSes except Windows and OpenBSD released patches for these covert channels.
In particular, macOS mitigated this issue by removing the searchable queue from its PRNG selection, abandoning the idea of reserving IPIDs for non-repetition altogether (see XNU \href{https://github.com/apple-oss-distributions/xnu/blob/xnu-8019.61.5/bsd/netinet/ip_id.c}{v8019.61.5}, which introduced the change, and \href{https://github.com/apple-oss-distributions/xnu/blob/xnu-8792.41.9/bsd/netinet/ip_id.c}{v8792.41.9}, which fixed a minor bug for avoiding IPID zero).

\subsection{Fragment Injections} \label{subsec:fraginject}

In a \textit{fragment injection}, the attacker plants one or more malicious fragments appearing to originate from some trusted source in a target machine's reassembly buffer.
If any of these fragments' IPIDs match that of a legitimate fragmented packet from the trusted source, a malicious fragment may replace a legitimate one during reassembly.
This technique has been used to poison DNS caches and forge domain validation certificates (Section~\ref{subsubsec:dnsattacks}), intercept traffic behind a NAT (Section~\ref{subsubsec:natintercept}), and shift time on NTP clients (Section~\ref{subsubsec:ntptime}).
The general template for fragment injection is:
\begin{enumerate}
    \item Ensure traffic from the trusted source to the target machine will be fragmented (e.g., by spoofing an ICMP ``Fragmentation Needed'' packet from the target to the source).

    \item Predict the IPID(s) that the trusted source will assign to the packets targeted for fragment injection, potentially using the probe comparisons outlined in Section~\ref{subsec:probecompare}.

    \item If needed, use path MTU discovery or other measurement methods to learn the necessary parameters for crafting malicious fragments.

    \item Spoof malicious fragments from the trusted source to the target machine.
\end{enumerate}
There is one salient necessary condition for successful fragment injections:
\begin{enumerate}[label=(FI\arabic*), ref=FI\arabic*]
    \item \label{cnd:fi:guess} An attacker must be able to guess, with reasonable probability, the IPID(s) that the trusted source will assign to the packets targeted for fragment injection.
\end{enumerate}

Fragment injections can be viewed as a special case of \textit{(de)fragmentation attacks}, exploits that abuse a machine's reassembly buffer size, fragment eviction policy, or reassembly strategy for unintended behavior.
For example, a denial of service attack known as ``teardrop'' abused the fact that older reassembly methods would fail if fragments overlapped by a few bits (\cve{CVE-1999-0015}).
Similar attacks were successful across every major OS at some point in time (e.g., \cve{CVE-1999-0052}, \cve{CVE-1999-0157}, \cve{CVE-1999-0431}, \cve{CVE-2000-0305}, and \cve{CVE-2004-0744}), and modern fragmentation attacks continue to be a concern for IP stack implementations (e.g., \cve{CVE-2020-3373}, \cve{CVE-2020-28041}, \cve{CVE-2021-3905}, and \cve{CVE-2023-24821}).
We focus specifically on fragment injections since these depend critically on IPID selection methods and their predictability.

\subsubsection{DNS Attacks} \label{subsubsec:dnsattacks}

\begin{figure}[t]
    \centering
    \includegraphics[width=0.7\textwidth]{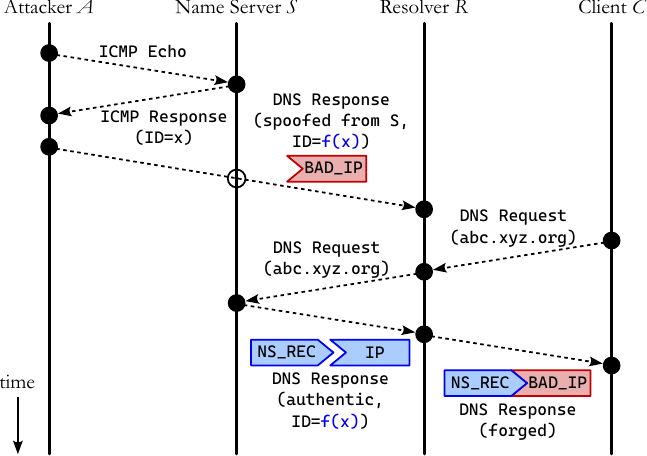}
    \caption{\textit{DNS Cache Poisoning Execution.}
    In this attack, an attacker $A$ is redirecting traffic for a specific domain (e.g., \texttt{abc.xyz.org}) to an IP address of its choosing.
    First, $A$ probes the DNS name server $S$'s current IPID, $x$.
    It then spoofs a DNS response fragment containing its chosen IP address to the DNS resolver $R$ that appears to have originated from $S$; critically, this fragment's IPID must match whatever IPID $f(x)$ that $S$ will use in its later DNS response.
    When $S$ responds to a DNS request for \texttt{abc.xyz.org}, its fragment containing the authentic IP address will be discarded in favor of the spoofed fragment with the same IPID already in $R$'s reassembly buffer, causing the reassembled DNS response to contain the attacker's chosen IP address.}
    \label{fig:dnspoison}
\end{figure}

In \textit{DNS cache poisoning}, an attacker redirects traffic to an IP address of its choosing by replacing a DNS resolver's cache entry for a valid domain name~\cite{Schuba1993-addressingweaknesses}.
The first cache poisoning attacks exploited the predictability of globally incrementing DNS transaction IDs (TXIDs) in the BIND protocol to trick DNS servers into accepting spoofed DNS replies~\cite{Arce1997-bindvulnerabilities}.
But forging entire DNS replies is not necessary for poisoning.
Instead, Herzberg and Shulman~\cite{Herzberg2013-fragmentationconsidered,Herzberg2013-vulnerabledelegation} (\figtext~\ref{fig:timeline}K) showed that poisoning can be achieved by placing spoofed fragments of a DNS response in a DNS resolver's reassembly buffer.
If any of these fragments' IPIDs match that of the real DNS response---which is easily guessed for globally incrementing or per-destination selection---a spoofed fragment will be reassembled with the rest of the DNS response, poisoning the cache (see \figtext~\ref{fig:dnspoison} for details).

Palmer and Somers~\cite{Palmer2019-firsttrydns} (\figtext~\ref{fig:timeline}T) later refined DNS cache poisoning to succeed on its ``first try''---i.e., with a single spoofed fragment---and Zheng et al.~\cite{Zheng2020-poisontroubled} (\figtext~\ref{fig:timeline}V) adapted it to work against DNS forwarders.
Klein~\cite{Klein2021-crosslayer} (\figtext~\ref{fig:timeline}X) revealed a weakness in the PRNG that Linux used to initialize its per-connection counters and generate per-bucket's stochastic increments, making DNS cache poisoning possible with per-connection and per-bucket IPIDs as well (see Linux commit \href{https://github.com/torvalds/linux/commit/f227e3ec3b5cad859ad15666874405e8c1bbc1d4}{f227e3}).
Finally, beyond using DNS cache poisoning to redirect traffic, Brandt et al.~\cite{Brandt2018-domainvalidation} (\figtext~\ref{fig:timeline}R) poisoned certificate authorities to obtain fraudulent certificates for arbitrary domains.

\subsubsection{NAT Packet Interception} \label{subsubsec:natintercept}

Gilad and Herzberg~\cite{Gilad2011-fragmentationconsidered,Gilad2013-fragmentationconsidered} (\figtext~\ref{fig:timeline}G) observed that when an external sender $S$ using per-destination IPIDs communicates with a NATted receiver, it treats the NAT device as its destination and thus draws from the same per-destination counter for all machines behind the same NAT.
Thus, if an off-path attacker $A$ controls one of the NATted machines $M$, it is easy to obtain the sender's per-destination IPIDs: $M$ simply pings $S$ and reports the reply's IPID to $A$.
Attacker $A$ also needs the destination port that the NAT uses to map traffic from $S$ to $M$, which $A$ obtains by spoofing packets from $S$ to the NAT with different guessed destination ports until one of them gets to $M$, who reports the correct destination port back to $A$.

Attacker $A$ can use these exfiltrated parameters to intercept traffic sent from $S$ to any receiver $R$ behind the NAT.
Suppose a legitimate packet sent by $S$ is fragmented as $(f_1, f_2)$, where $f_1$ contains the destination port for $R$ and $f_2$ contains the payload.
Using the predicted IPID, $A$ plants a spoofed second fragment $f_2'$ that is designed to reassemble with and discard the legitimate $f_1$ when it arrives.
It then spoofs a new first fragment $f_1'$ containing the destination port for $M$, the attacker-controlled NATted machine.
When the legitimate $f_2$ arrives, the NAT reassembles it with $f_1'$ and delivers it to $M$, completing the interception.
Generalized versions of this attack achieve off-path denials of service against inter-gateway tunnels when the attacker has puppet scripts behind both endpoints.

\subsubsection{Shifting NTP Time} \label{subsubsec:ntptime}

Malhotra et al.~\cite{Malhotra2016-attackingnetwork} (\figtext~\ref{fig:timeline}O) discovered an off-path attack leveraging predictable IPIDs to inject traffic into Network Time Protocol (NTP) client--server connections and shift time on NTP clients.
An attacker first spoofs an ICMP ``Fragmentation Needed'' packet from the target NTP client to its NTP server.
It then uses the puppet technique of Gilad and Herzberg~\cite{Gilad2011-fragmentationconsidered,Gilad2013-fragmentationconsidered} or the canary fragments technique of Knockel and Crandall~\cite{Knockel2014-countingpackets} to accurately predict the server's per-destination IPIDs.
This allows the attacker to spoof overlapping fragments from the server to the client containing phony timestamps which---depending on the client's reassembly policy for overlapping fragments---may get reassembled into the final packet.
This attack was never explicitly patched, though Malhotra et al.\ report that only $\approx 0.2\%$ of NTP servers and $\approx 1.3\%$ of tested NTP clients were vulnerable in 2016.
These numbers have likely only decreased as globally incrementing and per-destination IPIDs become less common.

\subsection{Measurements and Attacks Using IPv6 Fragmentation} \label{subsec:ipv6attacks}

Although our primary focus in this survey is IPv4's IPID, related vulnerabilities exist for IPv6.
Recall from Section~\ref{subsec:use} that fragmentation in IPv6 is performed only by senders and is achieved using the optional \textit{Fragment} extension header which includes a 32-bit identifier (i.e., the IPv6 IPID) for disambiguating packets during reassembly.
It was originally believed that IPv6 would obviate IPID-based exploits since (1) IPv6 packets only contain the \textit{Fragment} extension header and its constituent IPID when fragmentation is needed and (2) a combination of the IPv6 minimum link MTU and path MTU discovery would enable senders to resize their packets and avoid fragmentation a priori.
Unfortunately, as was the case for IPv4, path MTU discovery ``is not panacea''~\cite{Feng2022-pmtudnot}.
Off-path measurers or attackers can use a carefully crafted combination of ICMPv6 ``Echo Request'' and ``Packet Too Big'' messages specifying an MTU smaller than the IPv6 minimum to force IPv6 fragmentation~\cite{Morbitzer2013-tcpidle}. 
With this mechanism, sequences of IPv6 IPIDs can be probed and predicted to achieve off-path measurements and attacks analogous to those described in Sections~\ref{subsec:probecompare} and~\ref{subsec:fraginject}; some notable examples include TCP idle scans~\cite{Morbitzer2013-tcpidle}, IPv6 alias resolution~\cite{Beverly2013-ipv6alias}, and IPv6 router uptime characterization~\cite{Beverly2015-measuringcharacterizing}.
A recent study probing 20 million IPv6 addresses reports that 46.1\% of probed machines yield predictable IPv6 IPID sequences and may be vulnerable to IPID-based exploits~\cite{Huang2025-closerlook}.

\section{Comparative Analysis} \label{sec:comparison}

Section~\ref{sec:attacks} chronicled the co-evolution of IPID selection methods' OS implementations and the diverse measurements and attacks that exploited them.
In this section, we present a unifying mathematical model of IPID selection (Section~\ref{subsec:model}) and use it to compare and evaluate the methods' \textit{correctness} (Section~\ref{subsec:correct}), \textit{security} (Section~\ref{subsec:security}), and \textit{performance} (Section~\ref{subsec:performance}).
By parameterizing this comparison by a machine's expected rate of traffic, we reveal each selection method's relative (dis)advantages across a spectrum of use cases, summarized in Table~\ref{tab:comparison}.
These evaluations form the basis of our discussion and recommendations in Section~\ref{sec:recommend}.

\renewcommand{\arraystretch}{1.5}
\begin{table}[t]
    \caption{\textit{Summary of IPID Selection Method Comparisons.}
    The key comparisons and takeaways for each IPID selection method analyzed in Section~\ref{sec:comparison}.
    For brevity, we use \correcticon\ to indicate correctness, \secureicon\ for security, \timeicon\ for time complexity, and \memoryicon\ for space complexity.}
    \label{tab:comparison}
    \small{\begin{tabular}{p{1in}p{2.03in}p{2.03in}}
        \toprule
        \textbf{Selection Method} & \textbf{Strengths} & \textbf{Weaknesses} \\
        \midrule
        Globally Incrementing
            & Optimal \correcticon.
            Good \secureicon\ for sufficiently fast traffic, since rapid increments weaken correlation of probed IPIDs.
            Excellent \timeicon\ and \memoryicon, as atomic increments of a single 16-bit counter are fast and boast superior scalability under CPU contention.
            & Poor \secureicon\ for slow and moderate traffic, as infrequent, sequentially-incrementing counter values are very easily predicted. \\
        Per-Connection
            & Optimal \correcticon, \secureicon, and \timeicon, tying with or outperforming all other methods, though ``optimal \secureicon'' assumes off-path adversaries cannot probe connection counters.
            & Only usable by connection-bound traffic (incompatible with UDP, QUIC, etc.).
            Susceptible to downgrade attacks shifting to other methods.
            Unbounded \memoryicon. \\
        Per-Destination
            & Optimal \correcticon.
            Reasonable \memoryicon, obtained at the cost of extra mechanisms ensuring the number of destination counters does not grow too large (see weaknesses).
            & The worst \secureicon\ and \timeicon\ of all methods.
            Even if overall traffic is fast, traffic for any one destination may be slow, making its seqentially-incrementing counter easy to predict.
            Managing stale counters is time-intensive, and CPU contention over the destination hash table hurts scalability. \\
        Per-Bucket (Linux)
            & Optimal \correcticon, despite concerns about stochastic increments driving more frequent collisions.
            Near-optimal \secureicon\ for slow traffic, where large, noisy stochastic increments make prediction difficult.
            \timeicon\ Good scalability, though never as performant as globally incrementing.
            & Poor \secureicon\ for moderate and fast traffic, where counters are accessed so often that stochastic increments reduce to sequential increments, suffering similar predictability to per-destination.
            Reasonable \memoryicon\ that scales with user RAM, but often heavier than other methods. \\
        PRNG (searchable queue or iterated Knuth shuffle)
            & Worse \correcticon\ than counter-based methods, but reserving sufficiently many IPIDs for nonrepetition mitigates birthday paradox collision issues for all but the fastest traffic.
            Near-optimal \secureicon.
            Reasonable \memoryicon.
            & Reasonable \timeicon\ for single-CPU contexts, but scales very poorly due to contention over the searchable queue or cyclic permutation.
            Best for slow traffic settings. \\
        PRNG (pure)
            & Optimal \secureicon, producing IPIDs uniformly at random.
            \timeicon\ Fastest general-purpose method with trivial scalability (no CPU contention).
            \memoryicon\ Memoryless.
            & Collision probabilities follow the birthday paradox, yielding the worst \correcticon\ of any method.
            Usable only for slow traffic. \\
        \bottomrule
    \end{tabular}}
\end{table}
\renewcommand{\arraystretch}{1}

Several works surveyed in Section~\ref{sec:attacks} include formal modeling of IPID selection~\cite{Chen2005-exploitingipid,Herzberg2013-fragmentationconsidered,Ensafi2014-detectingintentional,Zhang2018-onisinferring,Klein2022-subvertingstateful}, but focus only on the exploit or target selection method at hand.
Similar modeling efforts exist for TCP sequence numbers, where security is analyzed only with respect to specific exploits (e.g.,~\cite{Qian2012-offpathtcp,Cao2016-offpathtcp,Medeiros2010-effectivetcp,Zhao2013-detectingcovert}) and performance analyses focus primarily on congestion control (e.g.,~\cite{Mathis1997-macroscopicbehavior,Padhye1998-modelingtcp,Cardwell2000-modelingtcp}).
Our goal, in contrast, is to evaluate all seven IPID selection methods on level footing, providing an intuitive synthesis for practitioners weighing the tradeoffs of different methods.

\subsection{Mathematical Model} \label{subsec:model}

We model a server $S$ communicating with one or more clients and analyze properties of the IPIDs it assigns to outgoing packets.
To capture the stochastic nature of network traffic, we assume $S$ sends packets with non-trivial IPIDs according to a Poisson process with rate $\lambda > 0$.
We define the ``unit time'' of this Poisson process as the average time for a packet sent by $S$ to be routed to its destination and have all its fragments either reassembled or evicted from the destination's reassembly buffer.\footnote{In practice, unit time can range from $\sim$10~ms (a fast ping) to 120~s (the longest recommended reassembly timeout~\cite{RFC791,RFC1122}). Assuming a unit time of 10~ms and an average packet size of \numprint{1500} bytes, 1~Kbps translates to $\lambda \approx 2^{-10.2}$, 1~Mbps translates to $\lambda \approx 2^{-0.3}$, 1~Gbps translates to $\lambda \approx 2^{9.7}$, 100~Gbps translates to $\lambda \approx 2^{16.3}$, and 1~Tbps translates to $\lambda \approx 2^{19.7}$.}
This time interval is important for two reasons.
First, it is the window of uncertainty for adversaries trying to correlate IPIDs obtained by probing $S$ (Section~\ref{subsec:probecompare}), since other packets that $S$ sends in the unit time between its reply to some probe and when that reply is received will further change its current IPID value.
Second, this is the interval during which fragments in the same reassembly buffer with the same IPIDs may interact, as required for fragment injections (Section~\ref{subsec:fraginject}).

Poisson processes are classically used in network traffic models that prioritize analytical elegance over high-fidelity representation and prediction.
Specifically, packet interarrival times are adequately captured by simple Poisson processes when aggregating large-scale traffic~\cite{Cao2003-internettraffic} or over short time scales~\cite{Paxson1995-widearea,Karagiannis2004-nonstationarypoisson}, though traffic patterns do exhibit long-range dependence, self-similarity, and other non-stochastic properties~\cite{Leland1995-selfsimilarnature,Karagiannis2004-longrangedependence}.
Alternative traffic models such as Markov Modulated Poisson Processes~\cite{Bhat1994-renewalapproximations,Bali2007-algorithmfitting} and autoregression models~\cite{Hosking1981-fractionaldifferencing,Paxson1995-widearea,Chen2000-trafficmodeling} can better capture these time-dependent properties, but are less amenable to the unifying comparative analysis we perform.

\subsection{Analyzing Correctness} \label{subsec:correct}

The intended function of IPIDs is to support unambiguous packet reassembly after IP fragmentation (see Section~\ref{subsec:use}), so any measure of an IPID selection method's correctness should capture its ability to assign distinct IPIDs to packets whose fragments would otherwise be indistinguishable---i.e., those with the same source and destination IP addresses, source and destination ports (if applicable), and protocol number.
Formally, we say the server $S$ produces a \textit{collision} at a client $C$ if two fragments in the reassembly buffer of $C$ belong to different packets sent by $S$ but have the same IPID and protocol number.
We analyze each selection method's correctness in terms of its \textit{worst-case probability of producing a collision}, establishing a basis for fair comparison across both counter-based and PRNG-based methods.

In the worst case, all packets sent by $S$ are sent to the same client $C$ via the same protocol and are fragmented in transit; thus, every packet has the potential to cause a collision.
Recall from Section~\ref{subsec:model} that we assume outgoing traffic from $S$ occurs according to a Poisson process with rate $\lambda > 0$.
Let $N$ be the corresponding random variable counting the number of packets ``simultaneously in transit'' from $S$ to $C$, i.e., those sent by $S$ that have not yet been reassembled or evicted by $C$.
Then $N$ is Poisson-distributed, characterized by probability mass function $\mathrm{pmf}(N, \lambda) = \lambda^Ne^{-\lambda}/N!$ and expected value $\E{N} = \lambda$.
Formally, the worst-case probability of $S$ producing a collision at $C$ as a function of $\lambda$ is
\begin{align} \label{eq:correct:base}
    \Pr{\text{collision}} &= \sum_{n = 1}^\infty \Pr{\text{collision} \mid N = n} \cdot \Pr{N = n} \nonumber \\
    &= \sum_{n = 1}^\infty \Pr{\text{collision} \mid N = n} \cdot \mathrm{pmf}(n, \lambda),
\end{align}
where $\Pr{\text{collision} \mid N = n}$ is the probability that any of the $n$ distinct packets simultaneously in transit from $S$ to $C$ are assigned the same IPID (which varies by IPID selection method) and $\Pr{N = n} = \mathrm{pmf}(n, \lambda)$ follows from $N$ being Poisson-distributed.

\begin{figure}[t]
    \centering
    \includegraphics[width=0.6\textwidth]{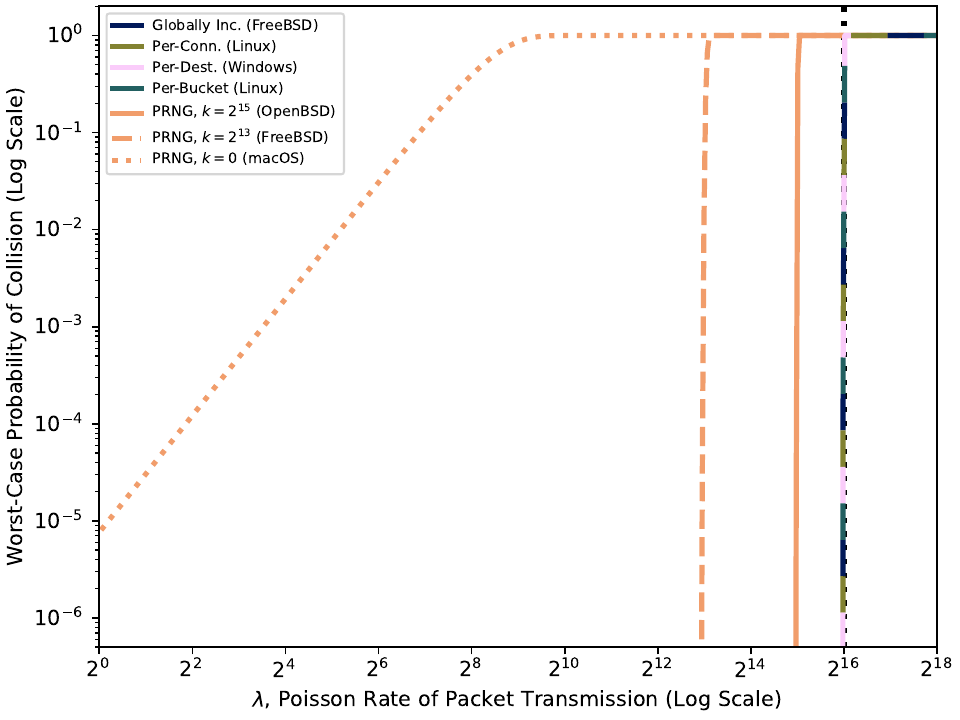}
    \caption{\textit{Comparison of IPID Selection Methods' Correctness.}
    Worst-case collision probabilities for globally incrementing (dark blue, \eqtext~\ref{eq:correct:global}), per-connection (olive, also \eqtext~\ref{eq:correct:global}), per-destination (pink, also \eqtext~\ref{eq:correct:global}), per-bucket (teal, simulation), and PRNG-based (orange, \eqtext~\ref{eq:correct:prngqueue}) IPID selection as a function of $\lambda$, the Poisson rate of packet transmission.
    The total number of distinct IPIDs, $2^{16}$, is shown as a black dotted line.
    Note that the collision probabilities of the four counter-based methods coincide.}
    \label{fig:correct}
\end{figure}

In Appendix~\ref{app:math:correct}, we derive these worst-case collision probabilities for all seven IPID selection methods.
The results are shown in \figtext~\ref{fig:correct}.
Globally incrementing, per-connection, and per-destination selection all increment sequentially, so their collision probabilities are effectively zero until there are so many packets simultaneously in transit that it is probable to exhaust all $2^{16}$ IPIDs and reuse at least one.
Perhaps more surprising is that per-bucket selection has nearly identical collision probabilities to the other counter-based methods even though its stochastic increments may skip some IPID values, cycle back around, and cause a collision in fewer than $2^{16}$ packets.
To understand why, recall that a per-bucket counter's increment value is chosen uniformly at random from 1 to the number of system ticks since the counter was last incremented (Section~\ref{subsec:implement}).
Any traffic rate $\lambda$ that generates enough packets to cause a probable collision also requests IPIDs so frequently that the per-bucket counter effectively increments sequentially.

PRNG-based methods yield non-negligible collision probabilities at slower traffic rates $\lambda$ than the counter-based methods do, with larger numbers $k$ of IPIDs stored for non-repetition corresponding to better collision avoidance.
Pure PRNG selection ($k = 0$) yields particularly poor correctness, succumbing to the birthday paradox: when $n$ packets (``people'') are assigned IPIDs uniformly at random from among the $2^{16}$ possible values (``birthdays''), it becomes quite likely that multiple packets have the same IPID.
At $\lambda = 2^5$ ($\approx 40~\text{Mbps}$), there is a 1\% chance of collision; at $\lambda = 2^7$ ($\approx 150~\text{Mbps}$), this probability increases to 10\%.
Methods that reserve $k \gg 0$ IPIDs for non-repetition achieve negligible collision probabilities until transmission rates become relatively fast ($\lambda \geq k$).

\subsection{Analyzing Security} \label{subsec:security}

Section~\ref{sec:attacks} categorized numerous IPID-based exploits as probe comparisons and fragment injections.
As highlighted by Conditions~\ref{cnd:pc:correlate} and~\ref{cnd:fi:guess}, the fundamental issue of \textit{predictability} in IPID generation underlies the success of both types of exploits.
With this motivation, we analyze each selection method's security in terms of its \textit{probability of an adversarial guess} which, informally, is the ability of an off-path adversary $A$ to guess the next IPID a server $S$ will assign in real time.

Formally, consider an off-path adversary $A$ trying to predict the next IPID generated by a particular IPID resource $i \in \{1, \ldots, r\}$ (i.e., a counter or PRNG) on $S$.
We assume $A$ knows---but cannot influence---both the rate $\lambda > 0$ of all outgoing traffic from $S$ and the rates $\lambda_i$ of only those packets whose IPIDs are assigned by resource $i$.
For globally incrementing and PRNG selection in which $r = 1$, we have $\lambda = \lambda_i$; for the selection methods with $r \geq 1$ counters, we have $\lambda \geq \lambda_i \geq 0$.
We further assume that $A$ can probe any IPID resource for its current IPID value except a per-connection counter, as no known exploits have demonstrated that this is possible for an off-path adversary.\footnote{If ever some future exploit enables an off-path adversary $A$ to probe per-connection counters, then the security of per-connection selection reduces to that of per-destination selection, as they both maintain multiple sequentially-incrementing counters (see Appendix~\ref{subsubsec:security:perconn} for details). Alternatively, if a downgrade attack~\cite{Feng2020-offpathtcp,Feng2022-offpathtcp,Feng2022-pmtudnot} forces $S$ to use some method other than per-connection selection, then its security reduces to that of the downgraded method.}

We allow $A$ a budget of $g \geq 1$ guesses and assume $A$ always guesses the $g$ maximum-likelihood IPIDs given the information available to it.
Thus, the probability of a successful adversarial guess is
\begin{equation} \label{eq:security:base}
    \Pr{\text{adv.\ guess}} = \max_{G \subset [2^{16}]~:~|G| = g} \Pr{\text{next IPID} \in G}
    = \max_{G \subset [2^{16}]~:~|G| = g} \left\{\sum_{x \in G} \Pr{\text{next IPID} = x}\right\},
\end{equation}
where $[2^{16}] = \{0, \ldots, 2^{16}-1\}$ denotes the set of all possible IPIDs, $G$ is a set of $g$ distinct IPIDs, and $\Pr{\text{next IPID} = x}$ is the probability that $x$ is the next IPID generated by resource $i$ (which varies among IPID selection methods).

\begin{figure}[t]
    \centering
    \includegraphics[width=\textwidth]{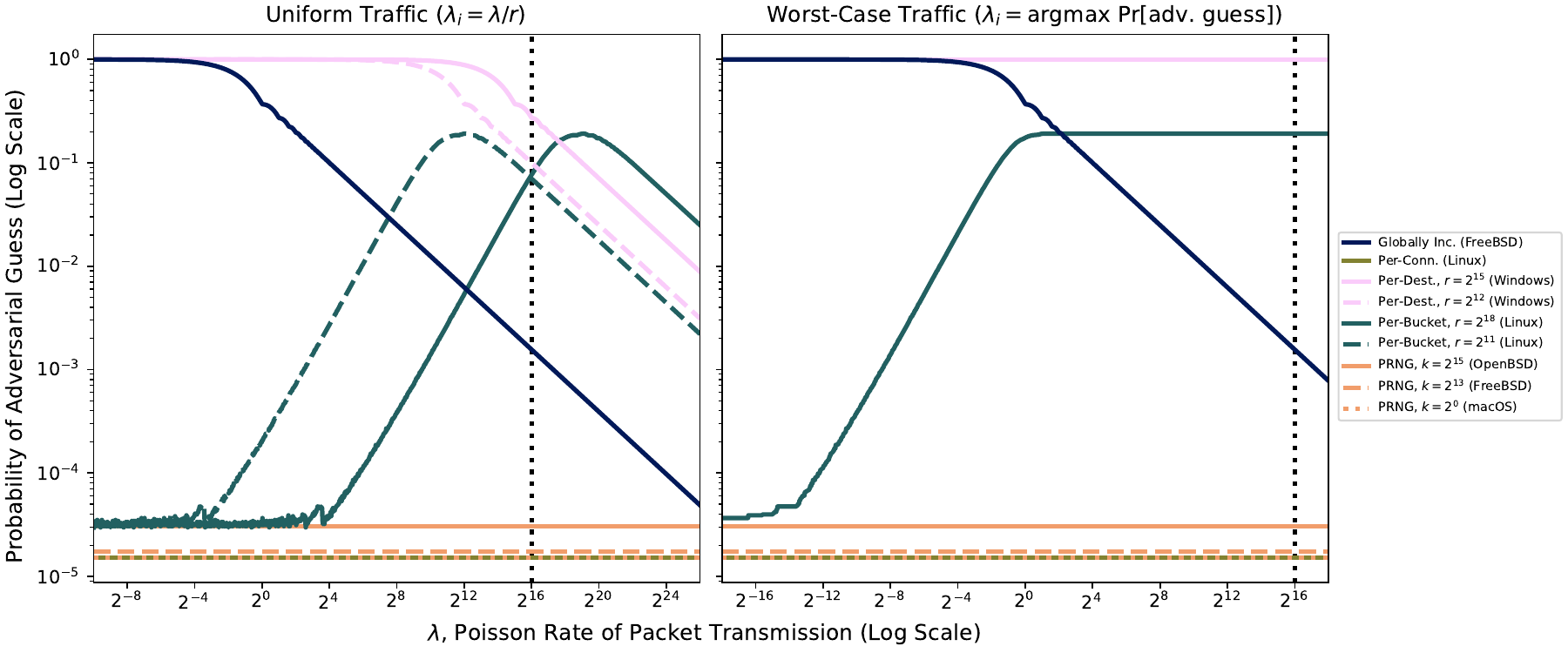}
    \caption{\textit{Comparison of IPID Selection Methods' Security.}
    Adversarial guess probabilities with $g = 1$ guess for globally incrementing (dark blue, \eqtext~\ref{eq:security:global}), per-connection (olive, \eqtext~\ref{eq:security:perconn}), per-destination (pink, also \eqtext~\ref{eq:security:global}), per-bucket (teal, \eqtext~\ref{eq:security:perbucket} with simulation), and PRNG-based (orange, \eqtext~\ref{eq:security:prng}) IPID selection as a function of $\lambda$, the Poisson rate of packet transmission.
    For selection methods that use $r \geq 1$ counters, we compare the case where traffic is uniformly distributed to each counter $i$ with rate $\lambda_i = \lambda / r$ (left) and the worst-case scenario when traffic is allocated to $\lambda_i$ such that the adversarial guess probability is maximized (right).
    The total number of distinct IPIDs, $2^{16}$, is shown as a black dotted line.
    An extended version of this figure showing $g = 10$ and $g = 100$ adversarial guesses is given in Appendix~\ref{app:math:security}, \figtext~\ref{fig:securityapp}.
    The absolute adversarial guess probabilities increase for larger $g$, but the comparisons among methods remain the same.}
    \label{fig:security}
\end{figure}

In Appendix~\ref{app:math:security}, we derive adversarial guess probabilities for the seven IPID selection methods as a function of $\lambda_i$.
To facilitate a fair comparison of results across selection methods, we visualize the results in two ways.
We first consider the ``uniform case'' where the $\lambda$-rate traffic is evenly distributed over all $r$ IPID resources of $S$, yielding $\lambda_i = \lambda / r$ (\figtext~\ref{fig:security}, left).
For per-connection, per-destination, and per-bucket selection where $r \geq 1$, this represents a baseline setting where all active connections or destinations generate roughly the same amount of traffic or where the bucket hash function distributes a similar amount of traffic to each bucket.
We then consider the ``worst case'' where $\lambda_i$ takes on whatever value is both feasible with respect to the total rate $\lambda$ and maximizes the adversarial guess probability (\figtext~\ref{fig:security}, right), i.e.,
\begin{equation} \label{eq:worstcaselambda}
    \lambda_i = \left\{ \begin{array}{cl}
        \lambda & \text{if $r = 1$}; \\
        \arg\max_{0 \leq \lambda_i \leq \lambda}\{\Pr{\text{adv.\ guess}}\} & \text{if $r > 1$}.
    \end{array} \right.
\end{equation}
These worst case $\lambda_i$'s capture the many ways $\lambda$ total traffic can be allocated among $r$ different IPID resources (e.g., bucket counters), surfacing the allocation that makes a given resource maximally predictable.
For example, is it easier to predict a given bucket counter $b$ when $b$ is responsible for nearly all $\lambda$ traffic, or when $b$ assigns very few IPIDs with the rest of the traffic being handled by other buckets?
Together, the uniform case emphasizes how the number of IPID resources impacts security while the worst case captures the fundamental predictability of any one IPID resource.

An IPID selection method is \textit{optimally secure} if its probability of an adversarial guess is exactly $g/2^{16}$, i.e., if the adversary can do no better than make its $g$ distinct guesses uniformly at random over all possible IPIDs.
Per-connection and pure PRNG selection are thus the only optimally secure methods, though per-connection's optimality depends on the potentially precarious assumption that an off-path adversary can never infer the status of another machine's connection counter or downgrade per-connection selection to some less secure method.
In contrast, the PRNG-based methods achieve (near-)optimal security based only on the assumption of a sufficiently random, correctly implemented, and well-seeded PRNG algorithm.

Our model confirms the probable success of idle scan attacks~\cite{Antirez1998-newtcp,Ensafi2014-detectingintentional}, predicting near-certain adversarial guesses for globally incrementing selection on any quiet channel (\figtext~\ref{fig:security}, $0 < \lambda < 1$).
It also confirms that Linux's addition of stochastic increments to per-bucket selection obfuscates otherwise predictable counter statuses in very quiet buckets ($\lambda < 2^{-12}$ for worst-case traffic; $\lambda < 2^{-1}$ for uniform traffic).
However, this obfuscation quickly dissipates as the rate of traffic increases.
Surprisingly, per-bucket selection is easier to predict than globally incrementing selection even at relatively low traffic rates ($\lambda > 2^2$ for worst-case traffic; $\lambda > 2^7$ for uniform traffic), revealing a subtle but significant security advantage to having all IPIDs assigned by a single resource.
As traffic rates increase, the number of increments to the single, global counter between any two of the adversary's probes also increases, weakening the correlation between subsequent probed IPIDs; at very fast rates of traffic ($\lambda > 2^{16}$), this correlation is so weak that globally incrementing selection approaches the unpredictability of PRNG-based methods.
Per-destination selection inherits the worst of both globally incrementing and per-bucket selection: it has multiple counters that always increment by one, rendering it the most predictable of all methods for all but the most extreme traffic rates.
Thus, like per-connection selection, its ``security'' lies not in its resilience to prediction but in the difficulty of accessing its counters, which has been subverted before.

\subsection{Analyzing Performance} \label{subsec:performance}

Our final dimension of comparison is performance, measured in both time and space complexity.
We perform this analysis in the context of multi-core servers where CPUs may contend over IPID resources.
To assess time complexity, we benchmarked each IPID selection method in ten independent trials per number of CPUs $c \in \{1, 2, \ldots, 128\}$ on a 128-core (dual-socket AMD EPYC 7713 Zen3), 512 GiB node on Arizona State University's Sol supercomputer~\cite{Jennewein2023-solsupercomputer}.\footnote{We repeated this benchmark on a 64-core (single-socket Intel Xeon Silver 4216), 256 GiB machine to validate our results are not architecture-dependent and obtained consistent results. See Appendix~\ref{app:benchmark:hardware}, \figtext~\ref{fig:timeintel} for details.}
In each trial, we count the number of IPIDs each of the $c$ CPUs assigns to packets in a 10 second period when concurrently and repeatedly scanning over a CAIDA trace of $\sim$29 million packets~\cite{CAIDA2019-caidaucsd}.
Details of our IPID selection method implementations are given in Appendix~\ref{app:benchmark:implementation}.

\begin{figure}[t]
    \centering
    \includegraphics[width=\textwidth]{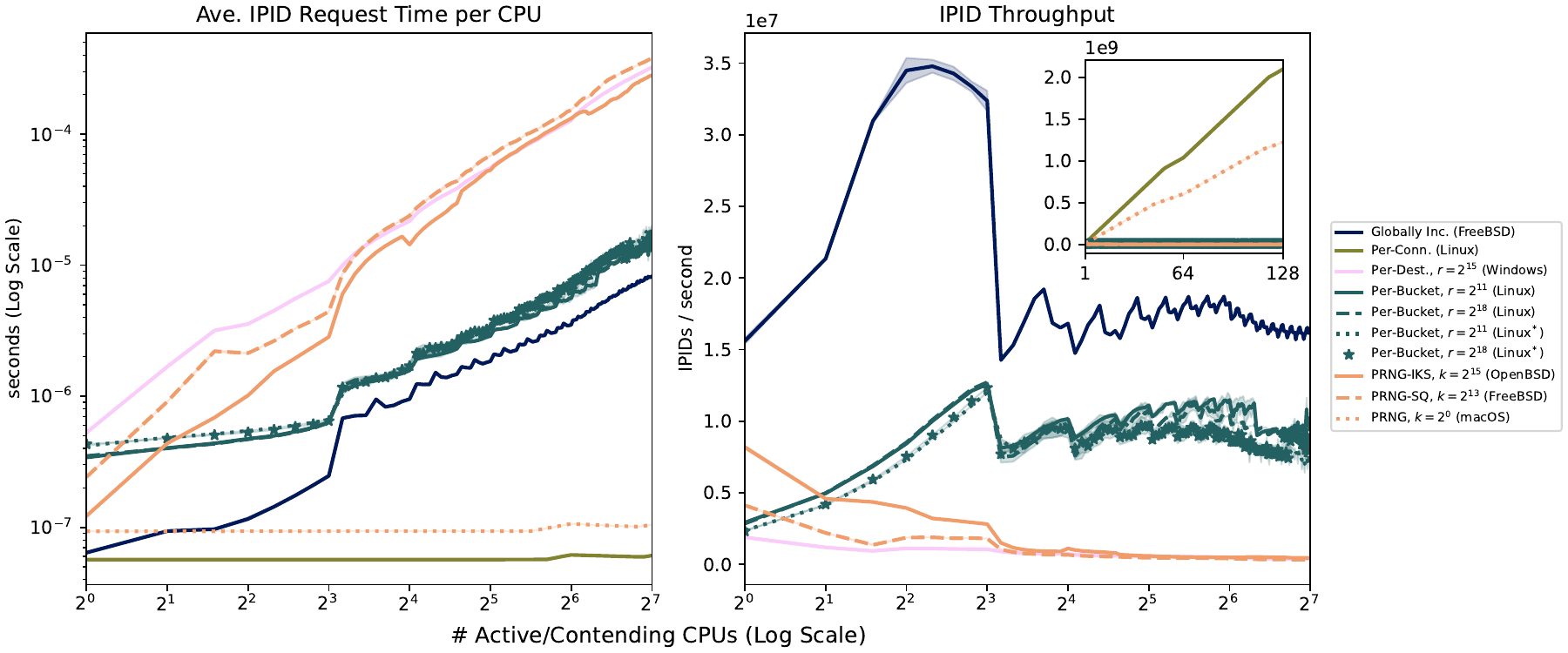}
    \caption{\textit{IPID Selection Methods' Time Complexities.}
    Average IPID request time per CPU (left) and IPID throughput per second for all CPUs combined (right, right inset) as a function of the number of CPUs concurrently assigning IPIDs via globally incrementing (dark blue), per-connection (olive), per-destination (pink), per-bucket (teal), and PRNG-based (orange) IPID selection.
    Per-bucket is implemented in two ways: those marked ``Linux'' (teal, solid and dashed) mimic Linux's actual implementation where bucket counters and last access timestamps are individually atomic but allow race conditions between them, while those marked ``Linux$^*$'' (teal, dotted and starred) maintain per-bucket locks.
    Values are shown as means (lines) and standard deviations (error tubes) over ten independent trials per IPID selection method and number of CPUs.}
    \label{fig:timeamd}
\end{figure}

The average IPID request times and packet throughput rates for each method are shown in \figtext~\ref{fig:timeamd}.
Unsurprisingly, per-connection selection is very fast (less than 0.1~$\mu$s per IPID request regardless of the number of CPUs).
When Linux gets IPIDs for connection-bound packets, it operates in the context of a socket data structure that at most one CPU is accessing per time.
Thus, obtaining an IPID is as simple as incrementing the socket's counter without any need to locate the counter via hashing or manage concurrency control.
Pure PRNG selection is similarly fast regardless of scale, since each CPU gets its own PRNG context.
The other methods require concurrency control to protect access to their shared IPID resources, causing their IPID request times to increase with the number of contending CPUs.
Among these methods, globally incrementing selection (implemented as a single atomic counter) is the fastest, even when all 128 CPUs contend over it.
Surprisingly, it outperforms even the per-bucket implementations whose primary reason for maintaining numerous buckets is to improve performance by reducing contention.
Globally incrementing selection is also desirably opportunistic, performing almost as well as per-connection selection in scenarios with very little contention ($1 \leq c \leq 4$).
Per-destination selection and PRNG-based methods that reserve IPIDs for non-repetition have poor performance and scalability, as their implementations rely on locking a single global resource (the hash table of destination counters, the searchable queue, or the shuffled permutation).

\begin{table}[t]
    \centering
    \caption{\textit{IPID Selection Methods' Space Complexities.}
    Where relevant, $r \geq 1$ is the number of IPID resources (i.e., counters or PRNGs), and $k$ is the number of IPIDs that PRNG-based methods reserve for non-repetition.
    $^*$The Windows implementation of per-destination selection technically has unbounded space complexity, but this table lists the expected memory usage for Windows Server v1904 (see text).}
    \label{tab:space}
    \begin{tabular}{llll}
        \toprule
        \textbf{Selection Method} & \textbf{\#16-bit Counters/IPIDs} & \textbf{Other Data} & \textbf{Memory} \\
        \midrule
        Globally Incrementing & 1 & N/A & 2 bytes \\
        Per-Connection & $r \geq 1$ active connections & N/A & Unbounded \\
        Per-Destination (Windows) & $r \leq 2^{15}$ active destinations$^*$ & $r$ 32-bit times & $\leq 192~\text{KiB}^*$ \\
        Per-Bucket (Linux) & $r \in [2^{11}, 2^{18}]$ buckets & $r$ 32-bit times & 12~\text{KiB}--1.5~\text{MiB} \\
        PRNG (searchable queue) & $k \in [2^{12}, 2^{15}]$ reserved IPIDs & $2^{16}$ lookup bits & 16--72~\text{KiB} \\
        PRNG (iterated shuffle) & $2^{16}$ IPIDs in shuffle & one 16-bit index & $\approx$ 128~KiB \\
        PRNG (pure) & N/A & N/A & None \\
        \bottomrule
    \end{tabular}
\end{table}

For space complexity, we consider all data structures that a selection method stores between IPID requests (see Table~\ref{tab:space}).
Globally incrementing selection stores a single 16-bit counter.
Per-connection and per-destination selection store one 16-bit counter per active connection or destination, respectively.
There are $\sim$3.7 billion unreserved IP addresses and thus there could be at most as many destination counters; the maximum number of connection counters is even larger since connections also consider source and destination ports and protocol numbers.
In practice, Windows fixes a purge threshold---$2^{12}$ destination counters in Windows 10 and $2^{15}$ in Windows Server---that puts a soft limit on the number of destination counters and 32-bit last access timestamps it stores before purging stale counters, though this limit can briefly be surpassed between purge sequences~\cite{Klein2022-subvertingstateful,Klein2024-privatecommunication}.
Linux treats connection counters as acceptably small members of its socket data structures, and implements per-bucket selection by storing one 16-bit counter and one 32-bit last access timestamp for each of its $2^{11}$ to $2^{18}$ buckets.
PRNG selection using a searchable queue of size $k \in [2^{12}, 2^{15}]$ stores the last $k$ IPID values generated by the PRNG, plus any data structure used to make searching fast; for example, FreeBSD and XNU search their queues in constant time by storing an array of $2^{16}$ bits where the $i$-th bit is 1 if and only if IPID $i$ is currently in the queue.
PRNG selection using the iterated Knuth shuffle stores the current permutation of all $2^{16}$ IPID values plus one 16-bit index for the next value in the permutation.
Finally, pure PRNG selection does not store any information between IPID requests other than the state of its PRNG.

Taken together, pure PRNG selection outperforms all but per-connection selection in terms of time and does not use any memory beyond its PRNG state; globally incrementing selection is the runner-up, outperforming the remaining methods while only using 2 bytes of memory.
Per-destination selection (with an upper limit on destination counters) and the non-repeating PRNG-based methods have moderate memory footprints, but their global locks are prohibitive performance bottlenecks that cannot be removed without threatening their correctness or security guarantees.
Finally, per-bucket selection achieves competitive IPID request times at the cost of potentially large (but bounded) memory usage.

\section{Recommendations} \label{sec:recommend}

Which IPID selection method is best?
As is the case for most worthwhile questions, it depends.
In this section, we synthesize insights from prior literature with the correctness, security, and performance evaluations of Section~\ref{sec:comparison} into concrete recommendations for IPID selection best practices.

Many have argued for avoiding fragmentation altogether by explicitly marking packets as atomic when possible~\cite{Kent1987-fragmentationconsidered,Bellovin1989-securityproblems,Gilad2011-fragmentationconsidered,Herzberg2013-fragmentationconsidered,Malhotra2016-attackingnetwork}, e.g., when packets are small or the path MTU is discovered as part of a TCP connection.
This best practice is already widely adopted.
All five OSes we reviewed make packets atomic when possible, and 82.4\% of packets in the CAIDA trace we used for benchmarking~\cite{CAIDA2019-caidaucsd} and an estimated 99\% of all network traffic is not fragmented~\cite{Shannon2002-folkloreobservations}.

\begin{recommend} \label{rec:atomic}
    Any IP packet that can avoid fragmentation (e.g., with a known path MTU) should be made atomic with $\text{DF} = \true$, $\text{MF} = \false$, and $\textit{Fragment Offset} = 0$.
\end{recommend}

But making packets atomic does not obviate the question of how their IPIDs should be assigned.
RFC 6864~\cite{RFC6864} states that an atomic packet's IPID can be any value and should be ignored by subsequent machines.
This latitude presents a previously unleveraged opportunity to secure not only the atomic packets---e.g., in the case of forced fragmentation (Section~\ref{subsec:fraginject})---but also the packets that may not be able to avoid fragmentation, such as DNS and QUIC traffic sent via UDP.
As we saw in Section~\ref{sec:comparison}, the security and performance of various IPID selection methods depend on the rate $\lambda$ of IPID assignment.
Choosing whether to assign IPIDs to atomic packets using a fixed value (as FreeBSD, macOS, and Linux do), a dedicated selection method (as Linux does), or the same selection method as for non-atomic packets (as OpenBSD does) changes those methods' effective IPID assignment rates, thus also changing their expected security and performance.

We propose a new approach to IPID selection based on user choice instead of OS developer hard-coding.
Our analyses in Section~\ref{sec:comparison} showed that although there is no universally ``best'' IPID selection method for every use case across all three dimensions of correctness, security, and performance, different methods achieve desirable tradeoffs for specific ranges of outgoing traffic.
Thus, in this new approach, OSes would implement several IPID selection methods and expose this choice as a network setting.
Users or system administrators would then estimate their rates of traffic to determine their use case and choose methods that best suit their needs, as we describe below.

\begin{recommend} \label{rec:config}
    OS implementations should make IPID selection configurable (e.g., as a network setting), enabling end users and system administrators to choose selection methods for their connection-bound and non-connection-bound packets.
    Per-connection selection for connection-bound traffic and per-bucket selection for all other traffic can be used as the default configuration.
\end{recommend}

\begin{figure}[t]
    \centering
    \includegraphics[width=0.6\textwidth]{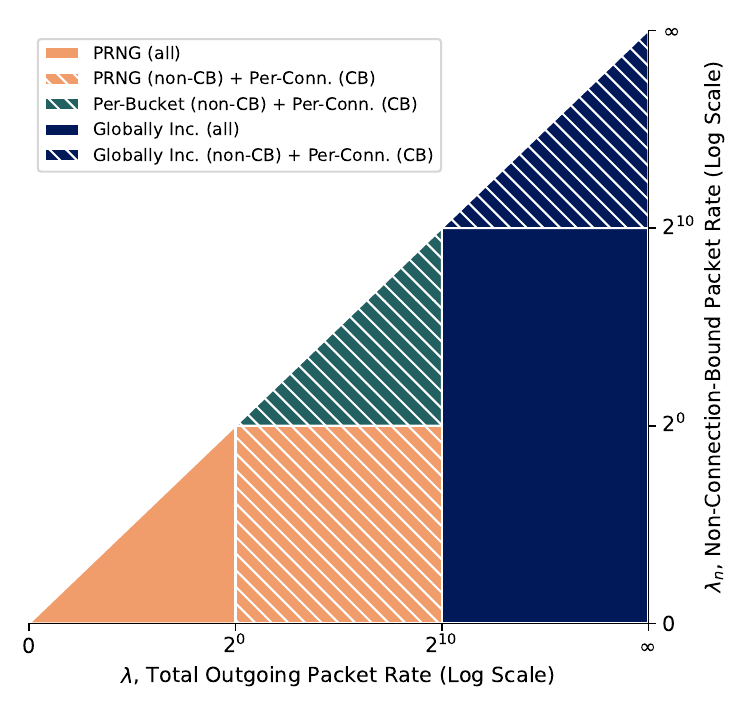}
    \caption{\textit{Recommended IPID Selection Method(s) by Use Case.}
    Depending on a machine's rates of total outgoing packets $\lambda$ and non-connection-bound (non-CB) packets $\lambda_n$, globally incrementing (dark blue), per-bucket (teal), or PRNG-based selection (orange) may provide the best correctness-security-performance tradeoff.
    Connection-bound (CB) packets can be handled separately with per-connection selection (white hatching) or combined with the rest depending on what provides the best tradeoffs.}
    \label{fig:phases}
\end{figure}

Suppose, then, that an end user or system administrator can estimate their Poisson outgoing packet rate as $\lambda = \lambda_c + \lambda_n$, where $\lambda_c$ is the Poisson rate of connection-bound packets and $\lambda_n$ is the Poisson rate of all other packets (i.e., non-atomic and non-connection-bound atomic packets).
We can use our evaluations from Section~\ref{sec:comparison} to partition the space of rates $(\lambda, \lambda_n)$ by use case and identify the best IPID selection method(s) for each.
See \figtext~\ref{fig:phases} for a visual summary.

\begin{recommend} \label{rec:phases}
    IPID selection method(s) should be chosen based on a machine's expected rates of total outgoing traffic and non-connection-bound outgoing traffic according to \figtext~\ref{fig:phases}.
\end{recommend}

In some of the use cases we describe below, it will be advantageous to assign IPIDs to connection-bound traffic separately from the rest (\figtext~\ref{fig:phases}, white hatching).
In these situations, connection-bound atomic packets should be assigned IPIDs using per-connection selection, storing connection counters in the corresponding socket data structures.
This is preferable to assigning them a fixed IPID value (e.g., zero) since it acts as a low-cost implementation of Postel's law---``be conservative in what you send, be liberal in what you accept''---hedging against forced fragmentation by a malicious adversary or other machines fragmenting packets downstream in violation of RFC 6848.

It remains to determine, for each use case, the best selection method for non-connection-bound packets and whether this method should also be used for connection-bound packets.
\begin{enumerate}
    \item \textit{Slow Overall} (e.g., printers, edge devices, IoT).
    If the total outgoing packet rate is slow ($\lambda_n \leq \lambda \leq 2^0 \approx$ 1~Mbps), our analysis shows that all methods are likely correct; per-bucket and PRNG-based selection achieve near-optimal security; and PRNG-based selection is faster than per-bucket selection as there is little to no CPU contention over IPID resources at these slow rates.
    Thus, PRNG-based selection can be used efficiently for all packets, regardless of whether they're connection-bound (\figtext~\ref{fig:phases}, solid orange).
    The specific choice of PRNG-based method can be left up to user preference.
    Pure PRNG selection has the best performance and security but much worse correctness; between the non-repeating methods, searchable queues use less memory while iterated Knuth shuffles have shorter request times, and larger numbers of reserved IPIDs $k$ achieve better correctness at the cost of slightly worse security.

    \item \textit{Slow Non-Connection-Bound, Moderate Connection-Bound} (e.g., home computers, HTTPS servers handling exclusively TCP traffic).
    If only the non-connection-bound traffic rate is slow ($\lambda_n \leq 2^0 \approx$ 1~Mbps) but the overall rate is too fast for pure PRNG selection to be correct and for non-repeating PRNG-based methods to be efficient ($\lambda > 2^0 \approx$ 1~Mbps), then the connection-bound traffic should be handled separately with per-connection selection so the remaining rate is slow enough to use a non-repeating PRNG-based method efficiently (\figtext~\ref{fig:phases}, hatched orange).
    This achieves near-optimal security for all traffic with high performance.

    \item \textit{Moderate Overall} (e.g., home computers, general purpose servers).
    In the region of moderate traffic rates ($2^0 < \lambda, \lambda_n < 2^{10} \approx$ 1~Gbps), pure PRNG selection's collision probabilities become unacceptably high and the performance bottleneck of per-destination and non-repeating PRNG-based selections' global locks cause them to scale poorly.
    Globally incrementing and per-bucket selection perform much better, and per-bucket's stochastic increments yield better security guarantees than global's sequential increments for these moderate rates.
    Those stochastic increments add the most noise for low traffic buckets, so connection-bound traffic should be handled separately with per-connection selection (\figtext~\ref{fig:phases}, hatched teal).

    \item \textit{Moderate Non-Connection-Bound, Fast Connection-Bound} (e.g., busy general purpose servers).
    In this use case, we reach perhaps the most surprising result of our analysis: reports of globally incrementing selection's disadvantages are greatly exaggerated.
    Globally incrementing selection not only scales better than all other methods under high contention (other than per-connection and pure PRNG), but also overtakes per-bucket selection in security (recall Section~\ref{subsec:security} for details).
    Thus, if the non-connection-bound packet rate is not fast enough to take advantage of globally incrementing selection alone ($\lambda_n < 2^{10} \approx$ 1~Gbps) but the total outgoing packet rate is ($\lambda \geq 2^{10} \approx$ 1~Gbps), then all traffic should be lumped together under globally incrementing selection (\figtext~\ref{fig:phases}, solid dark blue).

    \item \textit{Fast Non-Connection-Bound} (e.g., DNS servers).
    By the same logic as in the previous case, if the non-connection-bound packet rate is fast enough to benefit from globally incrementing selection on its own ($\lambda_n \geq 2^{10} \approx$ 1~Gbps), then it should do so.
    To avoid unnecessarily burdening the single global counter and its cache coherency protocol, the connection-bound traffic can use per-connection selection (\figtext~\ref{fig:phases}, hatched dark blue).
\end{enumerate}

\subsection{Implementation Suggestions} \label{subsec:osimplementation}

Altogether, we recommend that OSes implement four IPID selection methods: globally incrementing, per-connection, per-bucket, and a non-repeating PRNG-based method.
(Notably, per-destination selection is never recommended, owing to its poor performance and historically poor security regardless of use case.)
Against the benefits of secure and high-performance IPID selection configurable for each machine's specific use case, the costs of this proposed approach are small: some added code complexity and slight memory footprint inflation.
For example, Linux already implements both per-connection and per-bucket selection; adding globally incrementing selection and a non-repeating PRNG-based method would only use 16--128~KiB more memory.
FreeBSD also implements multiple methods (globally incrementing and PRNG with a searchable queue), using a user-configurable variable to choose between them.

At least in Unix-like OSes, there are standard configuration files and settings menus exposing networking variables to users.
For example, Linux's variable definitions in \texttt{/proc/sys/net/ipv4/} already enable runtime user configuration of various IPv4 fragmentation parameters, such as the reassembly timeout (\texttt{ipfrag\_time}).
To achieve our recommendations, users must be able to configure (1) which IPID selection method is used in general---choosing one of globally incrementing, per-bucket (default), or a non-repeating PRNG-based method---and (2) whether connection-bound traffic is handled separately by per-connection IPID selection (default: yes).
OS implementation can then follow Linux's method for differentiating connection-bound and non-connection-bound traffic and FreeBSD's method for activating the user-specified IPID selection method.
To aid users in choosing a suitable method for their use case, one could imagine a basic command line utility that leverages existing network monitoring tools (e.g., \texttt{iftop}) to estimate total and connection-bound outgoing traffic rates and then recommend a selection method according to \figtext~\ref{fig:phases}.

Our final recommendation is a word of caution.
In an effort to avoid creating new side channels when addressing old ones, we advise against dynamically updating IPID selection methods based on real-time traffic rates.
Recall that an off-path observer can easily infer which IPID selection method is active from observations of IPIDs over time~\cite{Bellovin1989-securityproblems,Mongkolluksamee2012-countingnatted}.
If the conditions for changing selection methods is known (as they would be for any open-source OS), any observable change in IPID patterns inherently leaks information about the status of those conditions, just as in hybrid leaks and downgrade attacks (Sections~\ref{subsubsec:bucketleaks}--\ref{subsubsec:tcpinject}).
If those conditions are further tied to something an attacker can influence (e.g., real-time traffic rates), the attacker may abuse this (e.g., by varying its probe rate) to force the target into using a selection method that it desires.
Instead, if different selection methods are preferable at different times (e.g., because traffic rates vary due to diurnal patterns or weekdays vs.\ weekends vs.\ holidays), OSes could support user-defined IPID selection method schedules.
This can only leak the existence of the schedule itself, but does not reveal other information about the state of the system and cannot be otherwise influenced by an off-path attacker.
In short, we advocate for user-configurable IPID selection methods, not methods that can be chosen or influenced by external actors beyond the user's control.

\subsection{Limitations} \label{subsec:limitations}

Our recommendations come with the following caveats and limitations.
First, as discussed in detail in Section~\ref{subsec:model}, our choice to model network traffic according to Poisson processes prioritizes unifying analytical elegance over total empirical fidelity.
This is a reasonable modeling choice for large-scale aggregated traffic or short time scales, but may not capture bursty, non-stochastic patterns.
In particular, our recommendation of using globally incrementing selection for scenarios with high traffic rates critically depends on continuous, rapid increments of the global counter; periods of slow traffic remain vulnerable to idle scans and other exploits that leverage easily predictable IPIDs.
A critical evaluation of our analytical results using simulated or deployed networking environments would be a valuable direction for future work.

\section{Conclusion} \label{sec:conclude}

In this survey, we collected a complete history of IPID-based exploits and the corresponding OS changes to IPID selection, categorizing these off-path measurements and attacks as either \textit{probe comparisons} aiming to infer information about other machines through changes in IPIDs over time or \textit{fragment injections} aiming to replace legitimate fragments with malicious ones during reassembly, causing packets to be poisoned or discarded.
We then presented the first comparative analysis of all seven IPID selection methods, formally analyzing their relative correctness and security and empirically evaluating their performance.
Of these evaluations, the most surprising is that globally incrementing selection---the first, simplest, and most dismissed of all IPID selection methods---is in fact the most collision-avoidant, secure, and performant choice for non-connection-bound packets when the rates of outgoing packets are very high.
Finally, we proposed a new approach to IPID selection that shifts the focus from developers making the ``best'' choice for their OS to users making the best choice for their use case across different ranges of total outgoing packet rates and non-connection-bound outgoing packet rates (see \figtext~\ref{fig:phases}).
It is our hope that further comparative analyses and theoretical evaluations of basic networking protocols will reveal similar best practices that can strengthen the correctness, security, and performance of all Internet-connected devices.

\section*{Software Artifacts}

Source code for the comparative analysis and performance benchmark in Section~\ref{sec:comparison} is available at
\ifanon
[URL omitted for blind review].
\else
\url{https://github.com/DaymudeLab/IPIDSurvey-Code}.
\fi

\begin{acks}
    We are deeply grateful to our anonymous reviewers and to Amit Klein for his detailed, constructive feedback and his elaboration on the current macOS and Windows implementations.
    J.J.D.\ is supported by the Momental Foundation's Mistletoe Research Fellowship, the ASU Biodesign Institute, and the National Science Foundation (CCF-2312537).
    A.M.E.\ is supported by DARPA (N6600120C4020).
    S.B.\ is supported by the ASU Fulton Fellowship.
    B.M--B.\ and J.R.C.\ are supported in part by the National Science Foundation (CNS-2141547).
\end{acks}

\bibliographystyle{ACM-Reference-Format}
\bibliography{ref}

\appendix

\section{Analytical Derivations} \label{app:math}

\subsection{Correctness Derivations} \label{app:math:correct}

To capture the stochastic nature of network traffic, we assume a server $S$ sends packets according to a Poisson process with rate $\lambda > 0$.
We further assume a worst-case scenario in which $S$ sends all its packets to the same client $C$ using the same protocol, every packet is fragmented, and the reassembly buffer of $C$ has unlimited size; thus, every packet has the potential to cause a collision.
Let $N$ be the corresponding random variable counting the number of packets ``simultaneously in transit'' from $S$ to $C$---i.e., those sent by $S$ whose fragments have not yet been reassembled or evicted by $C$---and let unit time be calibrated such that $\E{N} = \lambda$.
Then $N$ is Poisson-distributed, characterized by probability mass, cumulative distribution, and survival functions:
\begin{align}
    \mathrm{pmf}(N, \lambda) &= \frac{\lambda^Ne^{-\lambda}}{N!} \label{eq:poisson:pmf} \\
    \mathrm{cdf}(N, \lambda) &= \sum_{n=0}^N \mathrm{pmf}(n, \lambda) \label{eq:poisson:cdf} \\
    \mathrm{sf}(N, \lambda) &= \sum_{n = N+1}^\infty \mathrm{pmf}(n, \lambda) = 1 - \mathrm{cdf}(N, \lambda) \label{eq:poisson:sf}
\end{align}

\subsubsection{Globally Incrementing} \label{subsubsec:correct:global}

Globally incrementing IPIDs increment sequentially, so the only way for a collision to occur is if all possible IPID values are exhausted and at least one is reused.
Thus, the probability of $S$ producing a collision at $C$ among $N = n$ packets simultaneously in transit is
\begin{equation} \label{eq:correct:cond:global}
    \Pr{\text{collision} \mid N = n} = \left\{ \begin{array}{ll}
        0 & \text{if $n \leq 2^{16}$}; \\
        1 & \text{otherwise}.
    \end{array} \right.
\end{equation}
Substituting \eqtext~\ref{eq:correct:cond:global} into \eqtext~\ref{eq:correct:base} yields
\begin{equation} \label{eq:correct:global}
    \Pr{\text{collision}} = \sum_{n = 2^{16} + 1}^\infty \mathrm{pmf}(n, \lambda)
    = \mathrm{sf}(2^{16}, \lambda).
\end{equation}

\subsubsection{Per-Connection and Per-Destination} \label{subsubsec:correct:perconn}  \label{subsubsec:correct:perdest}

Per-connection and per-destination counters also increment sequentially and thus must exhaust all IPID values before a collision can occur.
Thus, \eqtext~\ref{eq:correct:global} also describes the probability of $S$ producing a collision at $C$ when using per-connection or per-destination selection.

\subsubsection{Per-Bucket (Linux)} \label{subsubsec:correct:perbucket}

All packets sent from $S$ to $C$ via the same protocol are hashed to the same bucket, so the total number of buckets has no bearing on the probability of collision.
However, bucket counters are not sequentially incrementing like those in the above methods.
Instead, Linux's stochastic increments make it possible to skip IPID values, cycle back around, and collide in fewer than $2^{16}$ packets.
Let $\Delta$ be a random variable representing the number of system ticks since $S$ last sent a packet; thus, the next bucket increment is chosen uniformly at random from $\{1, \ldots, \Delta\}$.
Since packet transmissions occur according to a Poisson process with rate $\lambda$, $\Delta$ is exponentially distributed with mean $t/\lambda$, where $t > 0$ is the number of system ticks per unit time.
For this analysis, we set $t = 3$, assuming that packets spend $\sim$10 ms in transit and there are $\sim$300 ticks/s.\footnote{The assumption of 300 system ticks/s is consistent with Arch Linux's implementation. Ubuntu is slightly slower, at 250 ticks/s. We chose Arch's larger value because it generates larger stochastic increments and thus higher probabilities of collision, suitable for a worst-case analysis.}

Unfortunately, even with these carefully constructed assumptions, $\Pr{\text{collision} \mid N = n}$ is challenging to bound analytically---let alone to calculate explicitly---because each event is dependent on all others.
Specifically, whether a new IPID produces a collision depends on the stochastic increments that produced all preceding IPIDs.
Thus, for the sake of comparison with other methods, we estimate these conditional probabilities by simulating the generation of IPIDs via stochastic increments and then computing the overall probability of collision using \eqtext~\ref{eq:correct:base}.\footnote{A careful reader might observe that the infinite sum in \eqtext~\ref{eq:correct:base} cannot be computed directly. Our simulations instead use the finite interval of $n \subset [0, \infty)$ containing all non-negligible Poisson probability mass; i.e., all $n$ such that $\mathrm{pmf}(n, \lambda) > \numprint{5e-324}$. We reuse this trick when evaluating \eqstext~\ref{eq:security:global} and~\ref{eq:security:perbucket}.}

\subsubsection{\texorpdfstring{PRNG (searchable queue of size $k$)}{PRNG (searchable queue of size k)}} \label{subsubsec:correct:prngqueue}

For the purposes of analysis, we assume that the PRNG algorithm in use sufficiently approximates a uniform distribution over all possible IPID values; i.e., any given IPID is chosen with probability $1/2^{16}$.
In the setting where no recent IPIDs are stored (i.e., when $k = 0$) and the number of packets simultaneously in transit is $N = n$, the probability of $S$ producing a collision at $C$ reduces to the birthday problem with $n$ ``people'' (the packets) and $2^{16}$ ``days'' (the IPIDs).\footnote{The OpenBSD, FreeBSD, and macOS implementations of PRNG selection reserve zero as a special IPID that is never returned (Section~\ref{subsec:implement}). This detail is omitted for clarity, but is easily addressed by replacing $2^{16}$ total IPIDs with $2^{16} - 1$ non-zero IPIDs.}
More generally, with a searchable queue of size $k \geq 0$, we are guaranteed that the next IPID will be distinct from the last $k$ generated IPIDs.
If $n > k$, the remaining $n - k$ IPIDs may collide with any newly generated IPID according to the birthday problem, yielding
\begin{equation} \label{eq:correct:prngqueuek}
    \Pr{\text{collision} \mid N = n} = \left\{ \begin{array}{ll}
        0 & \text{if $n \leq k$}; \\
        1 - \prod_{i=0}^{n-k-1}\left(1 - \frac{i}{2^{16} - k}\right) & \text{if $k < n \leq 2^{16}$}; \\
        1 & \text{if $n > 2^{16}$}.
    \end{array} \right.
\end{equation}
Substituting \eqtext~\ref{eq:correct:prngqueuek} into \eqtext~\ref{eq:correct:base} yields
\begin{align} \label{eq:correct:prngqueue}
    \Pr{\text{collision}} &= \sum_{n=k+1}^{2^{16}} \left(1 - \prod_{i=0}^{n-k-1}\left(1 - \frac{i}{2^{16} - k}\right)\right) \cdot \mathrm{pmf}(n, \lambda) + \sum_{n = 2^{16}+1}^\infty \mathrm{pmf}(n, \lambda) \nonumber \\
    &= \sum_{n=k+1}^{2^{16}} \left(1 - \prod_{i=0}^{n-k-1}\left(1 - \frac{i}{2^{16} - k}\right)\right) \cdot \mathrm{pmf}(n, \lambda) + \mathrm{sf}(2^{16}, \lambda).
\end{align}

\subsubsection{\texorpdfstring{PRNG (iterated Knuth shuffle reserving $k$ IPIDs)}{PRNG (iterated Knuth shuffle reserving k IPIDs)}} \label{subsubsec:correct:prngshuffle}

Supposing the permutation of all $2^{16}$ IPIDs is initialized uniformly at random and any returned IPID is swapped into a position chosen uniformly at random from among the $2^{16} - k$ previous positions in the permutation (including its own), the probability that any given IPID $x$ will be returned next is
\begin{equation} \label{eq:prngshuffle:prob}
    \Pr{\text{next IPID} = x} = \left\{\begin{array}{ll}
        0 & \text{if $x \in$ last $k$ IPIDs}; \\
        \frac{1}{2^{16} - k} & \text{otherwise}.
    \end{array}\right.
\end{equation}
This implies that the next IPID appears uniform at random over all IPIDs except the last $k$ returned.
So, just as for a searchable queue of size $k$, no IPID will collide with any of the last $k$ returned IPIDs and, if $n > k$, the remaining $n - k$ IPIDs may collide with any newly returned IPID according to the birthday problem.
Thus, \eqtext~\ref{eq:correct:prngqueue} also describes the collision probability for this method.

\subsubsection{PRNG (pure, no reserved IPIDs)} \label{subsubsec:correct:prngpure}

It is easily seen that pure PRNG selection---generating IPIDs uniformly at random over all $2^{16}$ possible values---is a special case of the previous two methods when $k = 0$.
Simplifying \eqtext~\ref{eq:correct:prngqueue} for this case yields
\begin{equation} \label{eq:correct:prngpure}
    \Pr{\text{collision}} = \sum_{n=1}^{2^{16}} \left(1 - \prod_{i=0}^{n-1}\left(1 - \frac{i}{2^{16}}\right)\right) \cdot \mathrm{pmf}(n, \lambda) + \mathrm{sf}(2^{16}, \lambda).
\end{equation}

\subsection{Security Derivations} \label{app:math:security}

Recall from Section~\ref{subsec:security} that we model an off-path adversary $A$ trying to predict the next IPID generated by a particular IPID resource $i \in \{1, \ldots, r\}$ (i.e., a counter or PRNG) on a server $S$.
Server $S$ sends packets according to a Poisson process with rate $\lambda > 0$ and its IPID resources $i$ assign IPIDs with rates $\lambda \geq \lambda_i \geq 0$ where $\sum_{i=1}^r \lambda_i = \lambda$.
Here, we analyze the probability of an adversarial guess against a particular resource $i$ as a function of $\lambda_i$.
Let $N_i$ be the Poisson-distributed random variable counting the number of packets $S$ sent in the last unit time whose IPIDs were assigned by resource $i$, and let time be calibrated such that $\E{N_i} = \lambda_i$.

\begin{figure}[t]
    \centering
    \includegraphics[width=\textwidth]{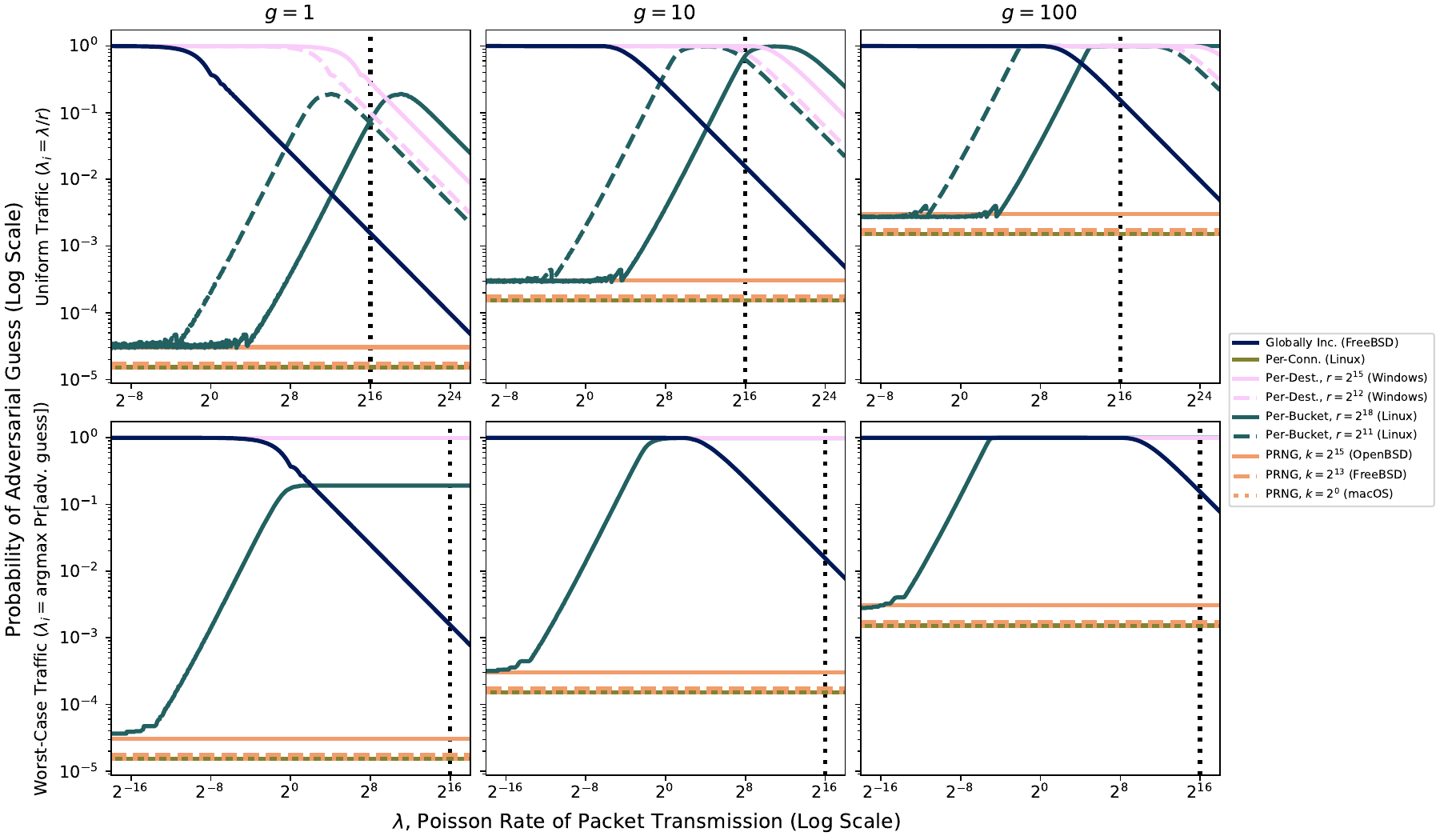}
    \caption{\textit{Comparison of IPID Selection Methods' Security with Multiple Adversarial Guesses.}
    Adversarial guess probabilities for $g \in \{1, 10, 100\}$ guesses for globally incrementing (dark blue, \eqtext~\ref{eq:security:global}), per-connection (olive, \eqtext~\ref{eq:security:perconn}), per-destination (pink, also \eqtext~\ref{eq:security:global}), per-bucket (teal, \eqtext~\ref{eq:security:perbucket} with simulation), and PRNG-based (orange, \eqtext~\ref{eq:security:prng}) IPID selection as a function of $\lambda$, the Poisson rate of packet transmission.
    For selection methods that use $r \geq 1$ counters, we compare the case where traffic is uniformly distributed to each counter $i$ with rate $\lambda_i = \lambda / r$ (top) and the worst-case scenario when traffic is allocated to $\lambda_i$ such that the adversarial guess probability is maximized (bottom).
    The total number of distinct IPIDs, $2^{16}$, is shown as a black dotted line.}
    \label{fig:securityapp}
\end{figure}

\subsubsection{Globally Incrementing} \label{subsubsec:security:global}

Globally incrementing selection uses one global counter for all IPID generation, so $r = 1$ and $\lambda_i = \lambda$.
The adversary $A$ can check the status of this counter by probing $S$ and examining the resulting IPID.
This tells $A$ the status of the global counter one unit time ago, when the reply was sent; w.l.o.g., suppose the IPID of this reply was 0.
Since the global counter increments once per packet sent by $S$ and there have been $N_i = N$ packets sent by $S$ in the last unit time, the current status of the global counter is $N \bmod 2^{16}$ and the next IPID will be $N+1 \bmod 2^{16}$.
Thus, the probability that the next IPID is $x$ is
\begin{align} \label{eq:security:next:global}
    \Pr{\text{next IPID} = x} &= \Pr{N+1 \equiv x \bmod 2^{16}} \nonumber \\
    &= \sum_{n=0}^{\infty} \Pr{N + 1 \equiv x \bmod 2^{16} \mid N = n} \cdot \Pr{N = n} \nonumber \\
    &= \sum_{n=0}^{\infty} \Pr{n + 1 \equiv x \bmod 2^{16}} \cdot \mathrm{pmf}(n, \lambda),
\end{align}
where $\Pr{N = n} = \mathrm{pmf}(n, \lambda)$ follows from the fact that $N$ is Poisson-distributed.
Substituting \eqtext~\ref{eq:security:next:global} into \eqtext~\ref{eq:security:base} yields
\begin{equation} \label{eq:security:global}
    \Pr{\text{adv.\ guess}} = \max_{G \subset [2^{16}]~:~|G| = g} \left\{\sum_{x \in G}\sum_{n=0}^{\infty} \Pr{n + 1 \equiv x \bmod 2^{16}} \cdot \mathrm{pmf}(n, \lambda)\right\}.
\end{equation}

\subsubsection{Per-Connection} \label{subsubsec:security:perconn}

Per-connection selection defines one randomly initialized counter $i$ per connection, which is specific to the IP addresses and ports of $S$ and its client $C$.
There are no known methods for the adversary $A \not\in \{S, C\}$ to obtain any information about the connection counter $S$ uses to assign IPIDs to packets sent to $C$.
Thus, regardless of the rate $\lambda_i$, adversary $A$ can do no better than make $g$ distinct guesses uniformly at random over all $2^{16}$ possible IPIDs:
\begin{equation} \label{eq:security:perconn}
    \Pr{\text{adv.\ guess}} = \max_{G \subset [2^{16}]~:~|G| = g} \left\{\sum_{x \in G} \Pr{\text{next IPID} = x}\right\}
    = \max_{G \subset [2^{16}]~:~|G| = g} \left\{\sum_{x \in G} \frac{1}{2^{16}}\right\} = \frac{g}{2^{16}}.
\end{equation}

Recall from Section~\ref{subsubsec:tcpinject} that downgrade attacks can force $S$ to use some selection method other than per-connection if one exists~\cite{Feng2020-offpathtcp,Feng2022-offpathtcp,Feng2022-pmtudnot}.
In this case, the probability of an adversarial guess reduces to that of the downgraded selection method.
Alternatively, if some future method enables $A$ to probe connection counters, the probability of an adversarial guess reduces to that of globally incrementing selection (\eqtext~\ref{eq:security:global}), but with $\lambda$ replaced by $\lambda_i$.

\subsubsection{Per-Destination} \label{subsubsec:security:perdest}

Per-destination selection uses one counter $i$ per destination IP address and protocol number.
As we reviewed in Sections~\ref{subsubsec:canaryfragments}, \ref{subsubsec:dnsattacks}, and~\ref{subsubsec:natintercept}, there have been many successful methods for probing destination counters~\cite{Gilad2011-fragmentationconsidered,Gilad2013-fragmentationconsidered,Herzberg2013-fragmentationconsidered,Herzberg2013-vulnerabledelegation,Knockel2014-countingpackets}, each specific to a particular OS implementation that was patched after disclosure.
For the purposes of this analysis, we assume the adversary can probe destination counters, focusing on the fundamental security of per-destination selection over the difficulty of accessing its counters.
The adversary can then use this information just as it did for globally incrementing selection to make inferences about the sequentially incrementing counter, so \eqtext~\ref{eq:security:global}---after replacing $\lambda$ with $\lambda_i$---also describes the probability of an adversarial guess for per-destination selection.
In \figstext~\ref{fig:security} and~\ref{fig:securityapp}, we plot per-destination's adversarial guess probabilities for $r = 2^{12}$ and $r = 2^{15}$ active destinations since these are the purge thresholds for Windows 10 and Windows Server, respectively~\cite{Klein2024-privatecommunication}.

\subsubsection{Per-Bucket (Linux)} \label{subsubsec:security:perbucket}

Per-bucket selection uses one counter per bucket.
We assume the adversary $A$ can probe any bucket counter $i$ on server $S$ that it desires by constructing a packet that hashes to that particular bucket.\footnote{Supposing that the bucket hash function appears to assign packets to buckets uniformly at random, the probability that at least one of an adversary's $a$ IPv4 addresses will hash into a target bucket among $r$ total buckets is $1 - (1 - 1/r)^a$. Even after Linux increased its maximum number of buckets to $r = \numprint{262144}$ (i.e., $2^{18}$)~\cite{Klein2022-subvertingstateful}, an adversary controlling $a = \numprint{10000}$ IPv4 addresses has a 3.74\% chance of finding a hash collision, $a = \numprint{100000}$ yields a 31.7\% chance, and $a = \numprint{1000000}$ yields a 97.8\% chance. These counts are similar to the sizes of large observed botnets. Machines with less RAM have fewer buckets, yielding higher hash collision probabilities at lower IP address counts.}
W.l.o.g., suppose the IPID of this reply is 0.
The bucket counter's current value depends on the number and timing of packets that hashed into this bucket in the unit time that elapsed since $S$ replied to $A$.

Recall that, for this method, $N_i \geq 0$ is the number of packets sent by $S$ with IPIDs assigned by bucket counter $i$ per unit time.
For each $j \in \{1, \ldots, N_i\}$, let $c_j$ be the increment value of bucket counter $i$ resulting in the IPID of the $j$-th packet sent after the reply to $A$.
Thus, the probability that the next IPID is $x$ is
\begin{align} \label{eq:security:next:perbucket}
    \Pr{\text{next IPID} = x} &= \Pr{\sum_{j=1}^{N_i+1} c_j \equiv x \bmod 2^{16}} \nonumber \\
    &= \sum_{n=0}^{\infty} \Pr{\sum_{j=1}^{N_i+1} c_j \equiv x \bmod 2^{16} \mid N_i = n} \cdot \Pr{N = n} \nonumber \\
    &= \sum_{n=0}^{\infty} \Pr{\sum_{j=1}^{n+1} c_j \equiv x \bmod 2^{16}} \cdot \mathrm{pmf}(n, \lambda_i),
\end{align}
where $\Pr{N_i = n} = \mathrm{pmf}(n, \lambda_i)$ follows from the fact that $N_i$ is Poisson-distributed.
Substituting \eqtext~\ref{eq:security:next:perbucket} into \eqtext~\ref{eq:security:base} yields
\begin{equation} \label{eq:security:perbucket}
    \Pr{\text{adv.\ guess}} = \max_{G \subset [2^{16}]~:~|G| = g} \left\{\sum_{x \in G}\sum_{n=0}^{\infty} \Pr{\sum_{j=1}^{n+1} c_j \equiv x \bmod 2^{16}} \cdot \mathrm{pmf}(n, \lambda_i)\right\}.
\end{equation}

Unlike globally incrementing selection where each $c_j = 1$, each per-bucket increment $c_j$ is chosen uniformly at random from a range $\{1, \ldots, \Delta_j\}$, where---as for our analysis of per-bucket collision probabilities in Section~\ref{subsubsec:correct:perbucket}---$\Delta_j$ is an exponentially distributed random variable with mean $t / \lambda_i$, where $t = 3$ is the number of system ticks per unit time.
This again makes formal analysis difficult; we thus use simulations to estimate $\Pr{\sum_{j=1}^{n+1} c_j \equiv x \bmod 2^{16}}$ via sampling and then compute the overall probability of an adversarial guess using \eqtext~\ref{eq:security:perbucket}.
In \figstext~\ref{fig:security} and~\ref{fig:securityapp}, we plot these adversarial guess probabilities for $r = 2^{11}$ and $r = 2^{18}$ buckets since these are Linux's minimum and maximum number of buckets, respectively.

\subsubsection{\texorpdfstring{PRNG (searchable queue of size $k$)}{PRNG (searchable queue of size k)}} \label{subsubsec:security:prngqueue}

We again assume (as in our correctness analysis in Section~\ref{subsubsec:correct:prngqueue}) that the PRNG algorithm in use sufficiently approximates a uniform distribution over all $2^{16}$ IPIDs.
We additionally assume that, in the worst case, the adversary knows the IPIDs of the last $k$ packets $S$ sent; i.e., it knows which IPIDs are in the searchable queue.
It is easy to see that, regardless of the rate $\lambda$, the probability that the next IPID is $x$ is
\begin{equation} \label{eq:security:next:prng}
    \Pr{\text{next IPID} = x} = \left\{ \begin{array}{ll}
        0 & \text{if $x \in$ queue}; \\
        \frac{1}{2^{16} - k} & \text{otherwise}.
    \end{array} \right.
\end{equation}
Substituting \eqtext~\ref{eq:security:next:prng} into \eqtext~\ref{eq:security:base}, we find that the adversary can do no better than make $g$ distinct guesses uniformly at random over the $2^{16} - k$ non-queued IPIDs:
\begin{equation} \label{eq:security:prng}
    \Pr{\text{adv.\ guess}} = \max_{G \subset [2^{16}]~:~|G| = g} \left\{\sum_{x \in G} \Pr{\text{next IPID} = x}\right\}
    = \min\left\{\frac{g}{2^{16} - k}, 1\right\}.
\end{equation}

\subsubsection{\texorpdfstring{PRNG (iterated Knuth shuffle reserving $k$ IPIDs)}{PRNG (iterated Knuth shuffle reserving k IPIDs)}} \label{subsubsec:security:prngshuffle}

We again assume the worst-case scenario that the adversary knows the IPIDs of the last $k$ packets that $S$ sent; i.e., it knows which IPIDs cannot be returned next.
In Section~\ref{subsubsec:correct:prngshuffle}, we showed that the next IPID generated by an iterated Knuth shuffle is equally likely to be any IPID except the last $k$ returned (\eqtext~\ref{eq:prngshuffle:prob}), implying that the adversary can do no better than make $g$ distinct guesses uniformly at random over the $2^{16} - k$ unreserved IPIDs.
Thus, \eqtext~\ref{eq:security:prng} also describes the probability of adversarial guess for this method.

\subsubsection{PRNG (pure, no reserved IPIDs)} \label{subsubsec:security:prngpure}

As in our correctness analysis (Section~\ref{subsubsec:correct:prngpure}), it is easily seen that pure PRNG selection is a special case of the other PRNG-based methods when $k = 0$.
Since every IPID is chosen uniformly at random from among all $2^{16}$ possible values, the adversary can do no better than guess at random.
Thus, \eqtext~\ref{eq:security:prng} (with $k = 0$) also describes the probability of adversarial guess for this method.

\section{Benchmark Details} \label{app:benchmark}

\subsection{Implementation Details} \label{app:benchmark:implementation}

Algorithm~\ref{alg:benchmark} details our implementations of the various IPID selection methods for the benchmarking results reported in Section~\ref{subsec:performance}.
We give some context for our design decisions below.

\begin{algorithm}[p]
    \caption{Benchmark Implementations of Multi-Core IPID Selection Methods}
    \label{alg:benchmark}
    \begin{algorithmic}[1]
        \State Let $c$ be a 16-bit atomic global counter. \label{alg:benchmark:globalstart}
        \Function{GloballyIncrementing}{ }
            \State Atomically fetch-and-add $v \gets c$ and $c \gets c + 1$.
            \State \Return $v + 1$. \label{alg:benchmark:globalend}
        \EndFunction

        \medskip

        \State Let $x$ be a local 16-bit counter. \label{alg:benchmark:connstart}
        Note that this estimates per-connection's performance, not its IPIDs.
        \Function{PerConnection}{ }
            \State Instantiate a counter $x$ and set $x \gets x + 1$.
            \State \Return $x$. \label{alg:benchmark:connend}
        \EndFunction

        \medskip

        \State Let $\ell$ be a lock, $h$ be a hash table of (16-bit destination counter, 32-bit last access time) pairs indexed by $(\text{src}_\text{ipaddr}, \text{dst}_\text{ipaddr})$ pairs, and $t_\text{purge}$ be the timestamp of the last purge sequence. \label{alg:benchmark:deststart}
        \Function{PerDestination}{$\text{src}_\text{ipaddr}$, $\text{dst}_\text{ipaddr}$}
            \State Lock $\ell$ and get the current time $t_\text{now}$.
            \If {$t_\text{now} - t_\text{purge} \geq 0.5~\text{s}$} initiate a purge sequence (see Section~\ref{subsec:implement}) and update $t_\text{purge} \gets t_\text{now}$.
            \EndIf
            \If {$(\text{src}_\text{ipaddr}, \text{dst}_\text{ipaddr}) \not\in h$} insert $h[(\text{src}_\text{ipaddr}, \text{dst}_\text{ipaddr})] \gets (\text{random IPID}, t_\text{now})$.
            \Else {} $h[(\text{src}_\text{ipaddr}, \text{dst}_\text{ipaddr})] \gets (\text{counter}(h[(\text{src}_\text{ipaddr}, \text{dst}_\text{ipaddr})]) + 1, t_\text{now})$.
            \EndIf
            \State Store $v \gets \text{counter}(h[(\text{src}_\text{ipaddr}, \text{dst}_\text{ipaddr})])$.
            \State Unlock $\ell$ and \Return $v$. \label{alg:benchmark:destend}
        \EndFunction

        \medskip

        \State Let $\ell$ be an array of $r$ bucket locks, $h$ be an array of $r$ 16-bit bucket counters, $t$ be an array of $r$ 32-bit last access times, and $key$ be a randomly generated 128-bit hash key. \label{alg:benchmark:bucketmutexstart}
        \Function{PerBucketMutex}{$\text{src}_\text{ipaddr}$, $\text{dst}_\text{ipaddr}$, $\text{prot\_num}$}
            \State Compute $j \gets $ \Call{SipHash}{$\text{dst}_\text{ipaddr}$, $\text{src}_\text{ipaddr}$, $\text{prot\_num}$, $key$} mod $r$.
            \State Lock $\ell[j]$.
            \State Get the current time $t_\text{now}$ and generate a random increment $inc \gets \mathcal{U}(1, \max\{1, t_\text{now} - t[j]\})$.
            \State Update $h[j] \gets h[j] + inc$ and $t[j] \gets t_\text{now}$.
            \State Store $v \gets h[j]$.
            \State Unlock $\ell[j]$ and \Return $v$. \label{alg:benchmark:bucketmutexend}
        \EndFunction

        \medskip

        \State Let $h$ be an array of $r$ 16-bit atomic bucket counters, $t$ be an array of $r$ 32-bit atomic last access times, and $key$ be a randomly generated 128-bit hash key. \label{alg:benchmark:bucketlinuxstart}
        \Function{PerBucketLinux}{$\text{src}_\text{ipaddr}$, $\text{dst}_\text{ipaddr}$, $\text{prot\_num}$}
            \State Compute $j \gets $ \Call{SipHash}{$\text{dst}_\text{ipaddr}$, $\text{src}_\text{ipaddr}$, $\text{prot\_num}$, $key$} mod $r$.
            \State Atomically swap the current time $t_\text{now} \to t[j]$ with the previous time $t_\text{old} \gets t[j]$.
            \State Generate a random increment $inc \gets \mathcal{U}(1, \max\{1, t_\text{now} - t_\text{old}\})$.
            \State Atomically fetch-and-add $v \gets h[j]$ and $h[j] \gets h[j] + inc$.
            \State \Return $v + inc$. \label{alg:benchmark:bucketlinuxend}
        \EndFunction

        \medskip

        \State Let $\ell$ be a lock, $q$ be a queue of the last $k$ IPIDs, and $m$ be an array of $2^{16}$ membership tracking bits. \label{alg:benchmark:prngqueuestart}
        \Function{PrngSearchableQueue}{ }
            \State Lock $\ell$.
            \State Repeatedly generate IPID values $v \gets \mathcal{U}(0, 2^{16} - 1)$ until $v \neq 0$ and $m[v] = \false$.
            \If {$|q| = k$} dequeue the last IPID $u$ from $q$ and set $m[u] = \false$.
            \EndIf
            \State Enqueue $v$ into $q$ and set $m[v] = \true$.
            \State Unlock $\ell$ and \Return $v$. \label{alg:benchmark:prngqueueend}
        \EndFunction

        \medskip

        \State Let $\ell$ be a lock, $p$ be the permutation of IPIDs, $i$ be the head index, and $k$ be the number of reserved IPIDs. \label{alg:benchmark:prngknuthstart}
        \Function{PrngKnuthShuffle}{ }
            \State Lock $\ell$.
            \State Let $v \gets p[i]$ be the next IPID in the permutation. \label{alg:benchmark:prngloop}
            \State Swap $p[i]$ with $p[i - \mathcal{U}(0, 2^{16} - k - 1) \bmod 2^{16}]$, then increment $i$.
            \If {$v = 0$} go to Line~\ref{alg:benchmark:prngloop}.
            \Else {} unlock $\ell$ and \Return $v$. \label{alg:benchmark:prngknuthend}
            \EndIf
        \EndFunction
    \end{algorithmic}
\end{algorithm}

\paragraph{Globally Incrementing}

We assume the single global counter shared by all CPUs is atomic, meaning it can be accessed and incremented by any one CPU without explicit locking.
Specifically, we use C++'s \texttt{std::atomic<uint16\_t>} which implements the necessary cache coherency protocols.

\paragraph{Per-Connection}

In practice, Linux handles TCP (connection-bound) traffic in sockets, and each socket data structure contains its own sequentially-incrementing counter.
Any time a thread is setting up a packet as part of a TCP connection, it's responding to a system call that already provided the socket structure as context.
So there's no cost to ``locating'' the connection counter via some kind of hash function, like there is in per-destination and per-bucket selection.
Moreover, there's essentially no situation in which multiple cores would be contending over the same socket simultaneously.
So from the perspective of our performance benchmark, the complexity of requesting a per-connection IPID is as simple as standing up a \texttt{uint16\_t} and incrementing it.

\paragraph{Per-Destination}

We model our per-destination implementation based on the Windows reverse-engineering efforts of Klein~\cite{Klein2022-subvertingstateful}, maintaining a hash table of (16-bit counter, 32-bit timestamp) pairs indexed by source and destination IP address pairs.
Following the ``purge sequences'' that Windows Server (v1904) uses to limit its hash table sizes, we fix a purge threshold of $r = 2^{15}$ entries~\cite{Klein2024-privatecommunication}.
Purge sequences are initiated and stale entries are removed as described in Section~\ref{subsec:implement}.
Because adding, removing, and updating destination counters are not thread-safe, we lock the entire hash table for each IPID request.
Klein does not report on how Windows handles concurrency~\cite{Klein2022-subvertingstateful}, and this appears to the be only solution that eliminates all race conditions.

\paragraph{Per-Bucket (Linux)}

For each of the $r$ buckets, we maintain a counter and a last access timestamp.
Following Linux's implementation, we locate the bucket for a given packet using the SipHash-2-4 hash function on the packet's destination and source IP addresses, the protocol number, and a randomly generated 128-bit hash key.
A stochastic increment is generated using the difference between the current time and the last access timestamp, after which the increment is applied and the timestamp is updated.
Because multiple packets may simultaneously hash to the same bucket, some concurrency control is needed.
Unlike in per-destination selection, however, multiple buckets can be accessed concurrently, avoiding the need for a global lock.
We benchmarked two versions of per-bucket's concurrency control: one which protects each bucket with a lock, and another that follows Linux's actual implementation in making each bucket's counter and last access timestamp individually atomic.
The latter allows a race condition where concurrent accesses to the same bucket could generate stochastic increments based on the same last access timestep and then atomically apply them to the same bucket counter.
This does not necessarily pose significant issues for correctness or security, but may change the statistical properties of the IPID sequences slightly.

\paragraph{PRNG-based Methods}

Our implementation of PRNG selection with a searchable queue directly follows FreeBSD's implementation, and our implementation of the iterated Knuth shuffle follows OpenBSD's.
Both of these methods coordinate multiple instructions that must be made atomically: e.g., testing membership of a new IPID in the queue and then enqueuing it, or swapping two IPIDs in the permutation before advancing the permutation's start index.
To eliminate race conditions, we lock access to the associated data structures for each IPID request.
Our implementation of pure PRNG selection follows the macOS/XNU implementation, first reducing a 64-bit salt to a 16-bit salt, and then applying this reduced salt to an IPID chosen uniformly at random from among all $2^{16}$ values.
Since this method does not keep state, concurrency issues can be sidestepped by giving each thread its own random number generator.

\subsection{Benchmark Hardware} \label{app:benchmark:hardware}

\begin{figure}[t]
    \centering
    \includegraphics[width=\textwidth]{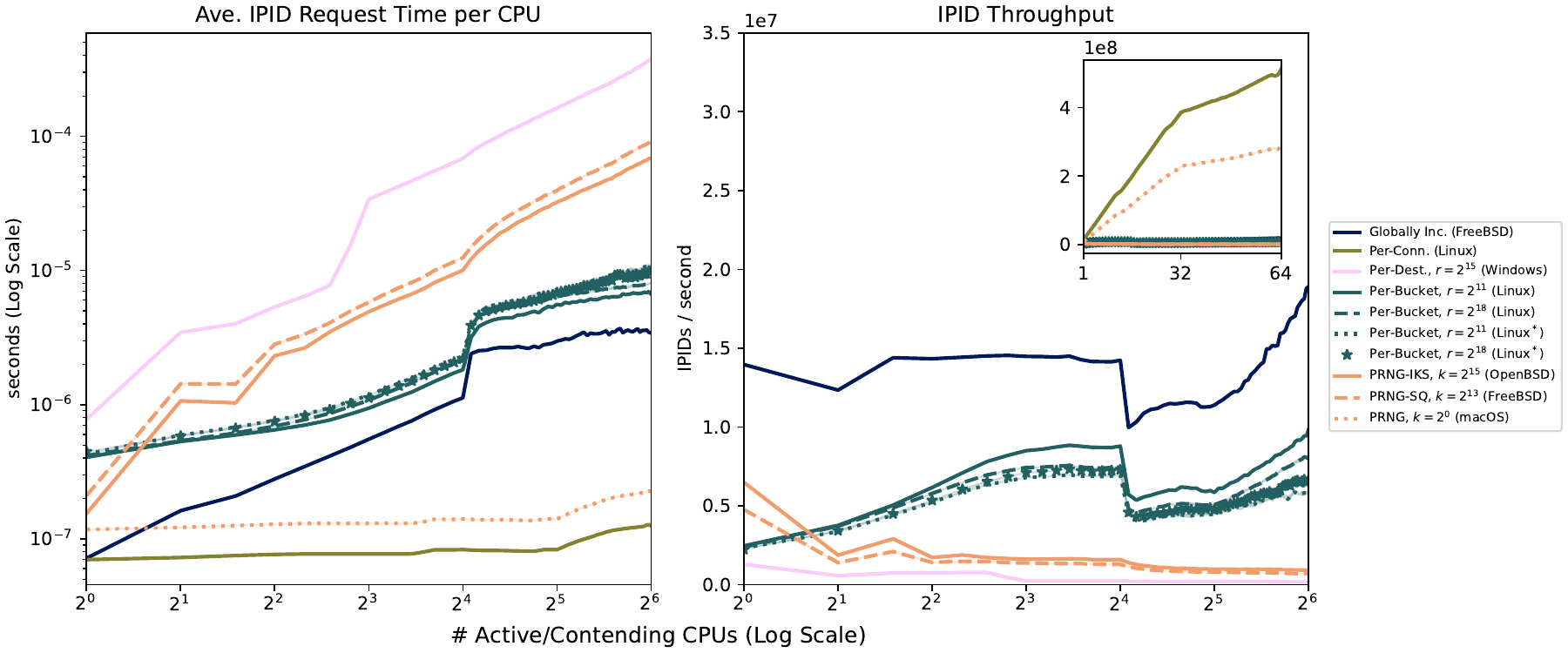}
    \caption{\textit{IPID Selection Methods' Time Complexities (Intel).}
    Analogous to \figtext~\ref{fig:timeamd}, but using a 64-core Intel CPU.}
    \label{fig:timeintel}
\end{figure}

As explained in Section~\ref{subsec:performance}, we ran our primary benchmark on a 128-core (dual-socket AMD EPYC 7713 Zen3), 512~GiB node on Arizona State University's Sol supercomputer~\cite{Jennewein2023-solsupercomputer}, yielding the results shown in \figtext~\ref{fig:timeamd}.
Since our benchmark depends heavily on CPU contention resolution mechanisms that may vary by CPU architecture, we ran a secondary benchmark on a 64-core (single-socket Intel Xeon Silver 4216), 256~GiB machine for the sake of comparison.
The results, shown in \figtext~\ref{fig:timeintel}, are consistent with those from the AMD CPU.
Per-connection and pure PRNG selection remain the highest-performing methods since they have no contention resolution to perform; among the remaining methods that perform contention resolution, globally incrementing selection is once again the best.
The only minor difference is that globally incrementing and per-bucket selection appear to scale slightly better on the Intel CPU than on the AMD CPU for 16 or more cores.

\end{document}